\documentclass[a4paper,11pt]{article}
\usepackage{graphicx}
\usepackage{grffile}
\usepackage{amsmath}
\usepackage{jcappub} 
\usepackage{hyperref}
\usepackage{xcolor}
\usepackage{lineno}

\newcommand\myadd[1]{{#1}}
\newcommand\myaddbegin{}
\newcommand\myaddend{}

\newcommand\ideal{idealized} 
\newcommand\Ideal{Idealized}

\newcommand\Omm{\Omega_{\text{M}}}
\newcommand\Omb{\Omega_{\text{b}}}
\newcommand\Oml{\Omega_{\Lambda}}
\newcommand\Omk{\Omega_{\text{K}}}
\newcommand\Ommz{{\Omega_{\text{M}}(z)}}
\newcommand\Omlz{{\Omega_{\Lambda}(z)}}
\newcommand\LCDM{$\Lambda$CDM}

\newcommand\hinvMpc{$h^{-1}$Mpc}

\newcommand\Hz{{H(z)}}
\newcommand\Hzsqr{{H^2(z)}}
\newcommand\Heff{{H_\text{eff}}}

\newcommand\Hfid{{H_\text{ref}}}
\newcommand\dep{\varepsilon}
\newcommand\deps{\varepsilon_s}
\newcommand\deplin{\varepsilon_\text{lin}}

\newcommand\tnow{t_\text{now}}
\newcommand\depinit{\varepsilon_\text{init}}
\newcommand\Rsmoo{R_\text{smoo}}
\newcommand\Rvoid{R_\text{void}}
\newcommand\Rmin{R_\text{min}}
\newcommand\Rmax{R_\text{max}}
\newcommand\relerr{e_\text{rel}}
\newcommand\lamlin{\lambda_\text{lin}}
\newcommand\sigR{\sigma_\text{R}}

\newcommand\colhead[1]{{#1}}

\arxivnumber{2407.03882} 
\title{Cosmology with voids}

\author[a]{Benjamin C. Bromley}
\author[b]{and Margaret J. Geller} 
\affiliation[a]{Department of Physics and Astronomy, University of Utah, 115 S 1400 E, Salt Lake City, UT, 84112, USA}
\affiliation[b]{Smithsonian Astrophysical Observatory, 60 Garden Street, Cambridge, MA 02138, USA} 

\emailAdd{bromley@physics.utah.edu}

\abstract{Voids are dominant features of the cosmic web. We revisit the cosmological information content of voids and connect void properties with the parameters of the background universe. We combine analytical results with a suite of large $n$-body realizations of large-scale structure in the quasilinear regime to measure  the central density and radial outflow of voids. These properties, estimated from multiple voids that span a range of redshifts, provide estimates of the Hubble parameter, $\Omm$ and $\Oml$. The analysis assumes access to the full phase-space distribution of mass within voids, a dataset that is not currently observable. The  observable properties of the largest void in the universe may also test models. The suite  of large $n$-body realizations enables construction of lightcones reaching $\sim$3,000~\hinvMpc. Based on these lightcones, we show that large voids similar to those observed are expected in the standard \LCDM\ model.
}

\begin{document}
\maketitle
\flushbottom

\section{Introduction}

Cosmic voids fill the universe. The vast, empty voids span tens of megaparsecs and contain few, if any, bright galaxies. Increasingly ambitious panoramic redshift surveys reveal the cosmic web with filaments, sheets and clusters of galaxies delineating a sea of voids, e.g., \cite{geller1989, SDSS2002, LCRS1996, 2dF2003, SDSSiii2012, GAMA2015, WiggleZ2018, HectoMAP2023, DESI2023}. 

Despite their low mass density, voids play a key role in shaping dense regions of the universe \cite{bertschinger1985, white1987, rg91, ceccarelli2013}. They grow relative to the cosmic expansion. As they grow, they push matter onto shells at their outer boundaries. Their growth also  feeds into galaxy clusters and superclusters. Like the massive, collapsed structures, voids reflect the ``background cosmology'', the mean density of matter and dark energy, and the overall expansion rate. They also contain the signature of the primordial  spectrum of density fluctuations, the seeds of the large-scale structure observed today. 

Early theoretical work considered void evolution \cite{alcock1979, aarseth1982, hoffman1982, bertschinger1985, melott1987, white1987, rg91, kauffmann1991, blumenthal1992}, while observational investigations \citep{elad1997, kauffmann1991, pan2011, sutter2012, micheletti2014, szapudi2015, finelli2016, mackenzie2017, kovacs2022, owusu2023, douglass2023} yielded void catalogs. More recent theoretical studies explored the void size distribution \cite{kamionkowski2009, sahlen2016, sahlen2018, correa2021, contarini2022, verza2024}, ellipticity distribution \cite{lavaux2010}, density and velocity profiles \cite{hamaus2014profile, hamaus2014, massara2018, stopyra2021, voivodic2020, voivodic2021, verza2022}, and redshift-space distortions \cite{hamaus2015, chuang2017, cai2016, nadathur2019lin}, including the Alcock-Paczynski (AP) test \cite{hamaus2022, radinovic2023} based on the expected statistical isotropy of large-scale structure. Other work emphasized the connection between void statistics and cosmological models \cite{spolyar2013, clampitt2013, cai2015, massara2015, zivick2015, voivodic2017, kreisch2019, perico2019, verza2019, contarini2021, bayer2021, kreisch2022, pelliciari2023, verza2023, wilson2023, vielzeuf2023}, and considered void properties in simulations \cite{nadathur2015a, nadathur2015b, nadathur2017, pollina2016, schuster2023, schuster2024}. Analyses of void data, based on redshift-space distortions and the AP test \cite{mao2017, hamaus2016, hamaus2017, hawken2017, achitouv2017, achitouv2019, nadathur2019, nadathur2020, aubert2022, hamaus2020, hawken2020}, along with the void size function \cite{contarini2023}, demonstrate the importance of voids as cosmological indicators.

Although linear perturbation theory for the growth of cosmic structure \cite{peebles1980} may describe the evolution of shallow voids on large scales, numerical simulations are required to connect theory and properties of voids as they emerge in the cosmic web. A challenge is that simulation volumes must be large with length scales of $\sim$1~Gpc to track ``typical'' superclusters and large voids. Most current hydrodynamical/$n$-body  simulations incorporate  cosmological parameters that best reflect the observations  \cite{springel2005, angulo2012, klypin2011, mccarthy2017, lee2021, kugel2023}. While not used for fitting cosmological parameters, these simulations provide model tests and may reveal tension between observations and theory.

Here we revisit the cosmological information contained in large voids. We confirm that void properties, including their size, interior mass density, and velocity flows, are sensitive to parameters of the background cosmology \cite{blumenthal1992}. Void properties alone may provide estimates of these parameters, drawn from a range of cosmological models with baryons, dark matter and dark energy. By connecting the properties of voids to the broader universe, we  test whether their dynamics are well-described by standard cosmology. Voids may eventually provide new insight into dark energy from rarefied environments where its dynamical effects are presumably greater than in the universe at large. 

We follow two complementary approaches to understand voids in the cosmological context. Following \cite{blumenthal1992}, we  examine the evolution of \ideal\ spherical voids numerically. This approach lays the foundation for relating the void depths and outflows to the underlying cosmology. We refine the cosmological constraints provided by the analytic analysis by generating a large suite of $n$-body models that treat the growth of large-scale structure in the quasilinear regime \cite{zeldovich1970, white2014, nadathur2019zel, stopyra2021, schuster2023}. This application of the adhesion approximation \cite{weinberg1990} enables the realization of snapshots of large-scale structure in patches of the universe along with the construction of lightcones that mimic large redshift surveys. We demonstrate that in theory, when the entire phase space is available, voids place tight constraints on the underlying cosmology. Any significant difference between these constraints from voids and other cosmological measurements may provide insight into the dynamical effects of dark energy.

We focus on the dynamics of matter flow \textit{within} voids where the ratio of dark matter density to dark energy density is generally low. Other work has considered the interface between (linear) voids and (nonlinear) neighboring high-density regions, including void shapes and void-galaxy cross correlations \cite{alcock1979, lee2009, hamaus2015, cai2016,  nadathur2019crossx, woodfinden2022}. These  analyses often depend on one or a few large-scale $n$-body simulations of the standard cosmology to compare theory and observation. Here we explore a wide range of models, sampling over 100 realizations of the cosmic density and velocity fields with $10^7$--$10^9$ particles to assess sensitivity to a broad range of cosmological parameters.
 
Section \ref{sec:bg} describes the suite of cosmological models we use as a test bed. In section~\ref{sec:voidsideal}, we examine \ideal\ voids to assess the role of matter and dark energy in void evolution. We use linear theory to interpret the results and we extract cosmological parameters from simple void properties (section~\ref{sec:cozestideal}). In section~\ref{sec:voidsrealish}, we use a fast $n$-body algorithm, the adhesion approximation \cite{weinberg1990}, to sample void properties in more realistic scenarios.  We assess these simulated voids as probes of the parameters of the background cosmology (section~\ref{sec:cosmoreal}). In section~\ref{sec:redshift} we consider voids in currently observable redshift space and highlight important differences between  the properties of voids in real and redshift space. We conclude in section~\ref{sec:conc}.

\section{Background cosmologies}\label{sec:bg}

To explore the cosmological information content of voids, we work with the standard parameters of the Friedmann-Lem\^{a}itre-Robertson-Walker (FLRW) framework. The overall expansion of the universe depends on  a matter density parameter $\Omm$, a parameter $\Oml$ associated with dark energy, and the Hubble parameter $H_0$. All three parameters refer to values at the present epoch (time $t = \tnow$). The mass density  $\Omm$ includes both baryons and cold dark matter; we approximate the effects of dark energy \cite{albrecht2006} with $\Oml$ as defined by a cosmological constant $\Lambda$. The set of parameters $\Omm$, $\Oml$ and $h$ --- $H_0$ in units of 100~km/s/Mpc --- define the ``background cosmology''. The cosmic web emerges as a pattern within this background cosmology. 

A scale factor, $a$, characterizes the evolution of the background cosmology. This quantity describes the time dependence of $\vec{r}$, the physical (``proper'') spatial separation between two points that comove with the cosmic expansion. In a coordinate frame tied to the expansion, the displacement between these two points, $\vec{x}$, is fixed; the scale factor provides the connection, with $\vec{r} = a\vec{x}$. Thus, $a$ encodes the details of cosmic expansion and evolves according to the Friedmann Equations,
\begin{gather}\label{eq:adot}
  {\dot{a}} = H_0 \sqrt{a^{-1}\Omm  + \Omk + a^2 \Oml},\\
  {\ddot{a}} = {H_0^2}\left[-a^{-2}\Omm/2 + a \Oml\right)
\end{gather}
where $\Omk$, the density parameter associated with the curvature of spacetime, is set by the requirement that $\Omm + \Oml + \Omk = 1$. The density parameters are directly related to other astrophysical quantities:
\begin{gather}
\label{eq:omm}
    \Omm = \frac{8\pi G}{3 H_0^2} \bar{\rho}_0, 
    \\
\label{eq:oml}
    \Oml = \frac{c^2}{3H_0^2}\Lambda ,  
\end{gather}
where $G$ is the gravitational constant and $\bar{\rho}_0$ is the present-day average density of the universe. We assume that contributions from relativistic particles in the universe are negligible.  

For reference, Table~\ref{tab:params} lists the main parameters and the symbols we adopt.

\begin{table}[htbp]
    \centering
    \small
\caption{List of parameters\label{tab:params}}
\begin{tabular}{lll}
\hline\hline
\colhead{symbol} & \colhead{parameter} & \colhead{note} 
\\
\hline
$a$ & cosmic scale factor & relates proper and comoving quantities \\ 
$c$ & speed of light \\
$D$ & density growth factor & linear theory parameter \\
$\delta$ & relative density contrast & $(\rho-1)/\bar{\rho}$ \\
$\dep$ & void depth & {${-\delta}$, averaged in the interior of a void} \\ 
\ &  \ &  $=$ $1$ for a vacuum, $0$ for mean cosmic density \\
$\deplin$ & void depth in linear theory & \  \\
$f$ & velocity growth factor & linear theory parameter ($\dot{D}/H$) \\
$G$ & gravitational constant & \ \\
$H$ & time-dependent Hubble parameter & $H_0$ denotes its present-day value \\ 
$h$ & present-day Hubble parameter & $H_0$ in units of 100~km/s/Mpc \\
$\Heff$ & effective $H$ & the relative flow rate within a void \\
$\Lambda$ & cosmological constant & a parameter in the FRWL model \\
$\lamlin$ & void linearity scale &  \\ 
$P$ & matter power spectrum & \ scaled with linear theory \\ 
$\rho$ & matter density & \ \\
$t$ & time & proper time in the background cosmology \\
$\tnow$ & proper time today & (age of the universe) \\
$z$ & redshift & $(1-a)/a$ \\
$\Omm$ & mass density parameter & includes both dark matter and baryons \\
$\Oml$ & dark energy density parameter & associated with $\Lambda$ \\
\hline
\end{tabular}
\end{table}

The evolution of the universe may be tracked either with proper time, $t$, or redshift $z$, the frequency shift of light observed from sources that comove with the expansion. In terms of the scale factor and model parameters, 
\begin{gather}
    t  =  \frac{1}{H_0} \int_0^a \left(\Omm \alpha^{-1}+\Omk+\Oml \alpha^2\right)^{-1/2} d\alpha\, , 
    \\
    z  =  \frac{1-a}{a}. 
\end{gather}
Although  we specify models with $\Omm$, $\Oml$ and $H_0$, all three present-epoch quantities also have redshift-dependent counterparts:
\begin{gather}
   \Ommz  =  (1+z)^3 \Omm H_0^2/\Hzsqr 
    \\
    \Omlz = \Oml H_0^2/ \Hzsqr 
    \\
    \Hz \equiv  {\dot{a}}/{a} = H_0 \sqrt{(1+z)^3\Omm + (1+z)^2 \Omk + \Oml}.
\end{gather}
These relationships enable  assessment of  the state of the background universe at early stages in its evolution. 

\subsection{Density fluctuations and the matter power spectrum}

Slight density fluctuations in the universe at early times were the seeds for the large-scale structure we observe today. The shallow troughs in the primordial matter density field were the progenitors of voids observed at later times. The matter power spectrum, $P(k)$, is a starting point for quantifying the properties of voids from the nascent matter density field.  The power spectrum is a function of  $k$,  the amplitude of a wavevector corresponding to a plane-wave Fourier mode of the matter distribution with wavelength $\lambda = 2\pi/k$. Formally, $P(k)$ is the power spectrum of the density contrast,
\begin{equation}\label{eq:delta}
    \delta(\vec{r}) = \frac{\rho(\vec{r})}{\overline{\rho}} -1,
\end{equation}
where $\rho$ is the local matter density and $\vec{r}$ is \myadd{a position vector} relative to some arbitrary origin. The power is the Fourier transform of the two-point correlation function of this density contrast and it measures the variance of mode amplitudes for each wavevector $\vec{k}$ when averaged within some large volume of the Universe. 

In the early expanding universe, self-gravity and the hydrodynamics of baryon-photon interactions mold primordial fluctuations from quantum effects.  The gravitational growth of small density fluctuations initially follows linear perturbation theory \cite{peebles1980},
\begin{equation}\label{eq:deltalin}
    \delta(\vec{x},t) = \frac{D(t)}{D(t_i)}\delta(\vec{x},t_i),
\end{equation}
where $D$ is the density growth factor from the growing-mode solution for $\delta$, and $t_i$ is some reference time. The dependence on comoving coordinates  emphasizes that, in linear theory, patterns of density fluctuations expand with the background universe. In terms of the scale factor, $a(t)$,
\begin{equation}\label{eq:D}
    D(a) = \frac{5}{2}\Omm H_0^2 H \int_0^a \dot{a}^{-3}da,
\end{equation} 
for cosmologies parameterized with a cosmological constant \cite{heath1977, percival2005}. Thus, given a primordial density field, we may scale the fluctuation mode amplitudes, and the power spectrum itself, to any epoch we choose. By convention, the primordial power spectrum refers to the epoch just after photons and baryons decouple in the expanding, cooling universe, but before density fluctuation mode amplitudes grow close to unity.  This primordial power is typically scaled to the present epoch.

The power spectrum depends sensitively on cosmology. However, remarkably few parameters set its shape and amplitude. These quantities include density parameters $\Omm$ and $\Oml$, along with the relative density of baryons, $\Omb$. Other ingredients include the Hubble parameter $H_0$ and $\sigma_8$, describing the amplitude of linear fluctuations of matter at a scale of 8~\hinvMpc. Formally, this observational quantity is connected to the power $P(k)$ through the convolution integral 
\begin{equation}\label{eq:sigR}
    \sigR = \frac{1}{(2\pi)^3} \int d\vec{k} P(k) W_{TH}^2(k,R) 
\end{equation}
where 
\begin{equation} 
    W_{TH}(k,R) =\frac{3}{(kR)^2}\left[\frac{\sin(kR)}{kR} - \cos(kR)\right]
\end{equation}
is the Fourier transform of a spherical tophat function of radius $R$, normalized so that it yields unity when integrated over space. Equation~\eqref{eq:sigR} gives the standard deviation of fluctuation amplitudes when sampled with a tophat window function of radius $R$; by convention, $R = 8$~\hinvMpc\  is chosen when quantifying the amplitude of fluctuations in a cosmological model. Other choices, for example $R = 50$~\hinvMpc, offer insight into the shape of the power spectrum beyond its normalization.

To calculate $P(k)$ for particular model parameters, we use the \texttt{CAMB} software package\footnote{The Python version of \texttt{CAMB} is at \href{https://github.com/cmbant/CAMB}{\texttt{github.com/cmbant/CAMB}}.} \cite{lewis2000}, based on the method of \cite{seljak1996} and the fast analytical prescriptions of \cite{eisenstein1998}, as implemented in the \texttt{nbodykit} package\footnote{The \texttt{nbodykit} package is available at \href{https://github.com/bccp/nbodykit}{\texttt{github.com/bccp/nbodykit}}.}. 

\subsection{Summary of model parameters}

We explore a range of cosmological parameters ($\Omm,\Oml,\Omk, h$) that bracket the \LCDM\ model (Cold Dark Matter plus a cosmological constant) favored by observations: $\Omm = 0.1$--$1$, $\Oml \leq (1-\Omm)$ and $h = 0.5$--0.9, with the curvature component set by the FLRW condition $1 = \Omm+\Oml+\Omk$. We highlight the  particular models  listed in table \ref{tab:cosmic} that include  an open CDM model with $(\Omm,\Oml,\Omk,h)=(0.3,0,0.7,0.6)$, a flat CDM model with $(\Omm,\Oml, \Omk, h)=(1,0,0,0.5)$, and a \LCDM\ model with $(\Omm,\Oml,\Omk, h)=(0.3,0.7,0,0.7)$. We include a density of baryons that is similar to values determined by observations \cite{planck2020} for models with $\Omm \geq 0.3$. The two low-density models (CDM-lite and $\Lambda$-lite) a significantly lower baryon density so that they are still dark-matter dominated. These models have similar ages (13--14~Gyr) and  comparable density fluctuation spectra. The models are illustrative; the analysis below does not depend sensitively on the particular selection.

\begin{table}[htbp]
    \centering
\caption{Cosmological model parameters\label{tab:cosmic}}
\begin{tabular}{lcccccccc}
\hline\hline
\colhead{name} & \colhead{$\Omm$} & \colhead{$\Oml$} & \colhead{$\Omb$} & 
\colhead{$\Omk$} & 
\colhead{$h$} & \colhead{$\sigma_8$} &   \colhead{$\sigma_{50}$} & \colhead{$\tnow$/Gyr}
\\
\hline
\LCDM &          0.3 & 0.7 & 0.044 & 0 & 0.7 & 0.8 & 0.048 & 13.47 \\
flat CDM &       1.0 & 0.0 & 0.044 & 0 &  0.5 & 1.2 & 0.033 & 13.04 \\
open CDM &       0.3 & 0.0 & 0.044 & 0.7 & 0.6 & 0.7 & 0.047 & 13.18 \\
CDM-lite &       0.1 & 0.0 & 0.004 & 0.9 & 0.65 & 0.3 & 0.035 & 13.51 \\
$\Lambda$-lite & 0.1 & 0.9 & 0.004 & 0 & 0.9 & 0.4 & 0.038 & 13.88 \\
\hline
\end{tabular}
\\[1.5pt]\parbox{5in}{\small 
The selected cosmological models are defined with $\Omm$, $\Oml$, and $\Omb$, as in the text, and $\Omk$ follows from $1 = \Omm+\Oml+\Omk$. The $\sigR$ columns refer to the standard deviation of the primordial matter density fluctuations averaged in tophat window functions of radius $R$ (equation~\eqref{eq:sigR}) evaluated at 8~\hinvMpc\ and 50~\hinvMpc. The final column is the age of each model, $\tnow$. All parameters  are evaluated at this age, including the fluctuation amplitudes.}
\end{table}

\begin{figure}[htbp]
\centering
\includegraphics[width=6in]{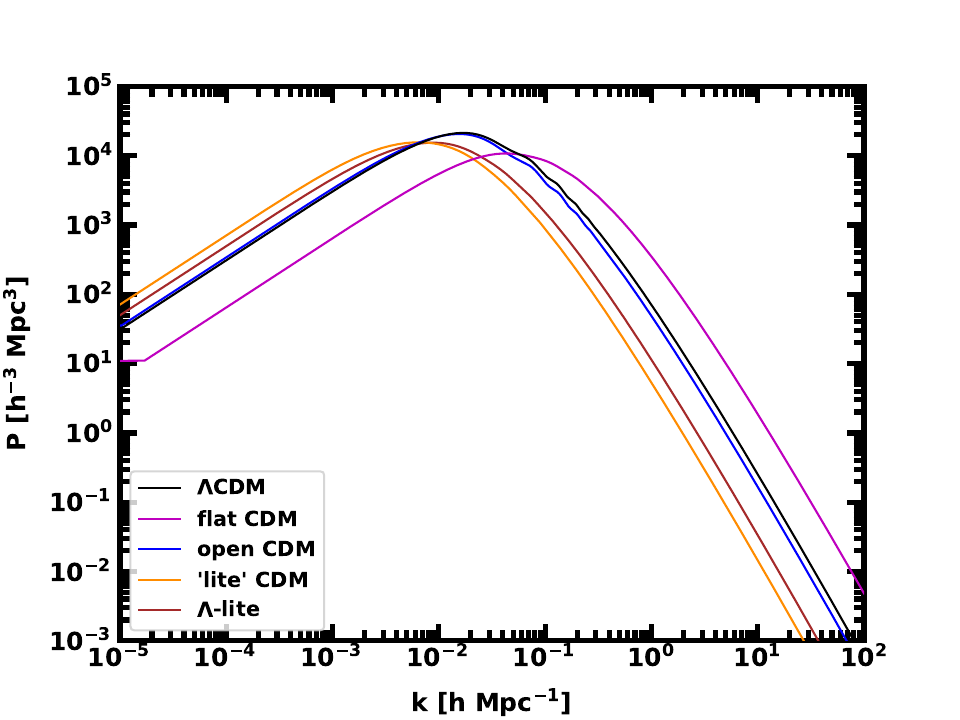}
    \caption{
    Linear matter power spectra of the models in table~\ref{tab:cosmic} normalized to the present epoch with the parameter $\sigma_8$. The Fourier transform of the tophat window function that defines $\sigma_8$ (light gray curve) highlights its sensitivity to the power in fluctuation modes. The \LCDM\ and open~CDM models are nearly identical, demonstrating the strong dependence of $P(k)$ on $\Omm$ as compared to $\Oml$ or $\Omk$.}\label{fig:powerspec}
\end{figure}

Figure~\ref{fig:powerspec} shows the power spectra associated with the models in table \ref{tab:cosmic}. The curves track the primordial power spectra scaled to the present day under the assumption that fluctuations grow according to linear theory. These curves enable assessment of the prevalence of voids --- or their progenitors --- as a function of  the void depth relative to the mean density of the universe. 

We next consider the dynamics of matter within voids in these illustrative models to assess the dependence of void dynamics on the background cosmological parameters. We then explore the inverse problem of inferring $\Omm$, $\Oml$ and $h$ from void properties. 

\section{\Ideal\ voids}\label{sec:voidsideal}

An informative starting point for exploring the connection between voids and their background cosmology is an idealized spherical region with a constant density that is, at some early time, slightly less than the average density in the universe. In isolation, this \ideal\ void evolves as if were its own miniature universe with a relatively small matter density and large Hubble parameter \cite{bertschinger1994, bertschinger1985}. When embedded in a uniform density background universe, we expect the void dynamics to be more complicated as material in the outermost regions of the more rapidly expanding void mixes with the background. Yet the bulk of the \ideal\ void interior evolves as a distinct FLRW world.

We begin this section by considering three approaches to tracking the evolution of \ideal\ voids. The first (section~\ref{sec:lintheory}) comes from linear perturbation theory \cite{peebles1980},  valid when the amplitude of the relative density contrast is small ($|\delta| \ll 1$). Because the average density contrast never exceeds unity within a void, linear theory offers a useful approximation to the evolution of the matter density there. The second approach (section~\ref{sec:za}) incorporates the Zel'dovich approximation \cite{zeldovich1970}, a surprisingly robust quasilinear method \cite{white2014} for tracking void evolution.  Finally, to approach calculating exact solutions, we use the ``shell model'' in reference~\cite{blumenthal1992} to numerically solve the equations of motion for mass elements in an \ideal\ void (section~\ref{sec:shell}). These numerical solutions provide an accurate accounting of the mass density even as the outer shells cross into the background universe. Following a brief comparison between these three approaches, we close this section with an assessment how the matter distribution within \ideal\ voids evolve in different background cosmologies. 

Throughout, we quantify our results with a void depth parameter \cite{blumenthal1992}
\begin{equation}\label{eq:dep}
\dep(r) = 1 - \frac{\left<\rho\right>_r}{\bar{\rho}}, 
\end{equation}
where $\left<\rho\right>_r$ is the average mass density within a radius $r$ and $\bar{\rho}$ is the mean cosmic density; within a tophat underdensity, $\dep$ is constant. In terms of the relative density contrast, $\dep = -\delta$ in an \ideal\ void, so that $\dep = 0$ indicates no underdensity; a value of unity indicates that the region contains no mass at all.

\subsection{Linear theory}\label{sec:lintheory}

Linear perturbation theory for the growth of cosmic structure \cite{peebles1980} describes the evolution of the mass density when relative fluctuations are small, with $|\delta| \ll 1$ (equation~\eqref{eq:delta}). Here, we provide the predictions of the theory.

In linear theory, the depth varies with redshift as
\begin{equation}\label{eq:deplin}
    \deplin(z)  = \frac{D(z)}{D(z_\text{init})}\dep(z_\text{init}), 
\end{equation}
where redshift $z_\text{init}$ is some initial (or reference) redshift, and the denisty growth factors $D$ are defined in  equation~\eqref{eq:D}. The theory also predicts flow rates for matter within a void. Including both the Hubble flow and the growing-mode solutions from linear theory, the proper radial velocity relative to the center of a void is \cite{peebles1980}
\begin{equation}\label{eq:vr}
v_r = \Hz r +  
    \frac{2 f g}{3 \Ommz \Hz},  
\end{equation}
where $r$ is the proper distance to the void center, $f$ is a velocity growth factor defined as
\begin{equation}\label{eq:f}
    f \equiv \frac{\dot{D}}{H} ,
\end{equation}
and $g$ is the peculiar acceleration. For a spherically symmetric void, 
\begin{eqnarray}\label{eq:g}
    g  &  =  & 
    -\frac{4\pi G \bar{\rho}}{r^2}\int_0^r \delta(s) s^2 ds 
    \\
    \label{eq:gideal}
    & = & +\frac{\Hzsqr \Ommz r}{2} \dep.
\end{eqnarray}
The void depth $\dep$ in the last equation refers to a tophat underdensity (equation~\eqref{eq:dep}) or the amplitude of the mean underdensity within comoving radius $x = r/a$. (Equivalent expressions, as in references \cite{hamaus2015, cai2015, nadathur2019crossx}, are often written in terms of $\Delta(x)$, the mean density contrast within comoving distance $x$.) 

An effective Hubble parameter, $\Heff$, describes the radial outflow rate of matter in the interior of an \ideal\ void. The linear theory prediction for $\Heff$ in units of the background Hubble expansion rate is \cite{peebles1980, kaiser1987}
\begin{eqnarray}
    \frac{\Heff}{H} & = &  \frac{v_r}{r \Hz} = 1 + \frac{2 f g}{3 \Ommz \Hzsqr r},   \\
    \label{eq:Hefflin}
    \  & = &  1 + \frac{f \dep}{3},
\end{eqnarray}
\myadd{where $r$ and $v_r$ are proper distance and speed, respectively. This expression}
follows from equations~\eqref{eq:vr} and \eqref{eq:gideal}. In section~\ref{sec:cozestideal} we show that this connection between the outflow rate, the growth factor, and void depth leads to a robust estimator of the background cosmology.

\subsection{The Zel'dovich approximation}\label{sec:za}

A related analytical approach for tracking void growth is the Zel'dovich approximation \cite{zeldovich1970, hidding2013, white2014}, which connects linear and nonlinear phases of structure formation. The Zel'dovich approximation quantifies the displacement of a point particle at some initial position by the peculiar gravity of density fluctuations in the approximation that these fluctuations grow according to linear perturbation theory. For example, a particle sitting at some location in the primordial density field  is accelerated by the gravity of fluctuations around it. Its subsequent motion is in the direction of that acceleration. At early times when linear theory applies, only the magnitude of the acceleration changes. Thus the motion of the particle is simply along a single displacement vector.  The approximation is formally dependent on linear theory, but in a Lagrangian or ballistic framework that empowers it beyond linear theory \cite{white2014}. 

The Zel'dovich approximation gives the displaced position $\vec{x}$ of a particle that is initially at rest at comoving location $\vec{q}$ as
\begin{equation}
    \vec{x} = \vec{q} + \vec{\psi},
\end{equation}
where the displacement vector $\vec{\psi}$ is
\begin{equation}
    \vec{\psi}(\vec{x}) \sim \vec{\nabla} \phi;
\end{equation}
here, $\phi$ is the peculiar gravitational potential, related to the density contrast $\delta(\vec{x})$ (equation~\eqref{eq:delta}) through the Poisson equation. In the Fourier domain, this expression is simpler \cite{white2014}:
\begin{equation}
    \vec{\psi} =\frac{1}{(2\pi)^3} \int d^3k \frac{i\vec{k}}{k^2} \delta(\vec{k}) \exp^{i\vec{q}\cdot\vec{k}},
\end{equation}
where $\delta(\vec{k})$ refers to Fourier modes of the density contrast. In the approximation, these modes all grow at the same rate (equation~\eqref{eq:D}), and the scaling of their collective amplitudes determines the time when the displacement is evaluated. For example, when $\delta$ is scaled to the present epoch, as for the primordial power spectrum in figure~\ref{fig:powerspec},  $\vec{x}$ gives the location of the particle at the present epoch. 

The particle's peculiar velocity --- describing motion that is distinct from the background expansion --- follows from the time derivative of the displacement:
\begin{equation}
    \dot{\vec{x}} = H f \vec{\psi},
\end{equation}
where $H$ is the Hubble parameter and $f$ is the velocity growth factor from linear theory (equation~\eqref{eq:f}).

The Zel'dovich approximation is commonly applied  in $n$-body codes where particles are placed on a regular grid and density fluctuations are chosen at random based on some primordial power spectrum. The approximation provides displacements and velocities up to a redshift where linear theory applies, and even beyond, so long as particle trajectories do not cross. Because the computational complexity of the Zel'dovich approximation is  set by a 3-D Fourier transform, the early stages of evolution in an $n$-body simulation are evaluated rapidly in a single step, without the more intensive calculations of an ordinary differential equation (ODE) solver. The ODE solver takes over only after linear theory breaks down.

Because of its extraordinary effectiveness beyond the linear regime, the ``quasilinear'' Zel'dovich approximation works well for voids \cite{yoshisato2006}, a conclusion that has been compellingly validated for voids as small as 10~\hinvMpc\ with numerical simulation, e.g., \cite{schuster2023}.  

Here, we implement the Zel'dovich approximation in both 1-D and 3-D codes to demonstrate that it describes void dynamics even when linear perturbation theory breaks down. Section~\ref{sec:compare} 
below
contains our results, comparing linear theory, the Zel'dovich approximation, and a nonlinear model for \ideal\
 void evolution, described next. 

\subsection{Nonlinear theory: The shell model}\label{sec:shell}

To determine the evolution of voids with depths that are of order unity, formally beyond the reach of linear perturbation theory and even the Zel'dovich approximation, we perform nonlinear calculations that more accurately track the flow of mass.  Following the strategy in reference~\cite{blumenthal1992}, we model a void as a series of thin, concentric shells, each with an assigned mass uniformly distributed within it. Newtonian mechanics provides the framework for solving the detailed evolution of these shells over distances that are much smaller than the Hubble length, $c/H_0 \approx 4,000$~Mpc. 

In our implementation, we adopt initial conditions as in \cite{blumenthal1992}, with a starting redshift of $z_i \gg 1$, and a spherical, constant-density region that is underdense relative to the rest of the universe by some small factor, $\depinit$, at that redshift.
The initial mass density is 
\begin{equation}\label{eq:rhoi}
\rho_i(x) = \bar{\rho}\left[1+\delta(x) \right] = \frac{3 H_0 \Omm}{8\pi G a_i^3}  
\times 
\begin{cases}
    1 - \depinit & \text{if}\ x < x_v 
    \\
    1 & \text{otherwise}
\end{cases}
\end{equation}
where $x$ is a comoving distance to the origin at the center of the underdense region, $x_v$ is the region's initial comoving radius, and $a_i = 1/(1+z_i)$ is the initial scale factor.  We choose a starting redshift that is roughly consistent with the epoch of recombination, $z_i \sim 1,000$,  to connect with the amplitude of the initial underdensities and fluctuations observed in the  Cosmic Microwave Background (CMB). We use a fixed comoving size ($x_v$) of the underdense region to more clearly compare void evolution in different cosmological models. 

For the initial velocities, we assume that matter closely tracks the peculiar flow of growing density perturbations as predicted by linear theory \cite{peebles1980}. We set the initial proper radial velocity of each shell according to equation~\eqref{eq:vr}, which gives the outflow speed as a function of distance from the void center. Since shells are very thin compared with the void radius, we just use the average of the shell's inner and outer radius for the distance.

With these starting conditions, individual mass elements evolve according to the following equation of motion, expressed in terms of the proper distance $r$ to the center of the region:
\begin{equation}\label{eq:ddotr}
    \ddot{r} =  -\frac{3 H_0^2\Omm} {2 r^2} \left\{\int_0^x 
    [1 + \delta(y)] y^2 dy\right\} + H_0^2 \Oml r,
\end{equation}
which follows from a Newtonian representation of cosmic acceleration, taking full advantage of spherical symmetry \cite{peebles1980}. As in \cite{blumenthal1992}, we choose mass elements as thin, concentric shells centered on the underdense region. Each shell contains mass in accordance with its initial volume and local mass density (equation~\eqref{eq:rhoi}).  The radial velocity comes from equation~\eqref{eq:vr}. We track the trajectories of these shells by solving the equation of motion (equation~\eqref{eq:ddotr}) numerically with Python's \texttt{solve\_ivp} routine in the \texttt{SciPy.integrate} module. 

In practice, we use a simple equation of motion where the gravitational acceleration on a shell at radius $r$ is $-GM(r)/r^2$, where $M(r)$ is the sum of the masses of shells interior to $r$ (Python's \texttt{numpy.cumsum} routine). Shell crossings can affect this summation and the overall evolution of the system. 

There are other approaches to tracking void evolution in the \ideal\ case, taking full advantage of the spherical symmetry and the assumption of uniform density within a void \cite{shesh2004, pace2010, wagner2015, dai2015, desjacques2018}. The shell model is straightforward to implement and has flexibility to manage more complicated density profiles, modified force laws, and any shell crossings.

\begin{figure}[htbp]
    \centering
    \includegraphics[width=6in]{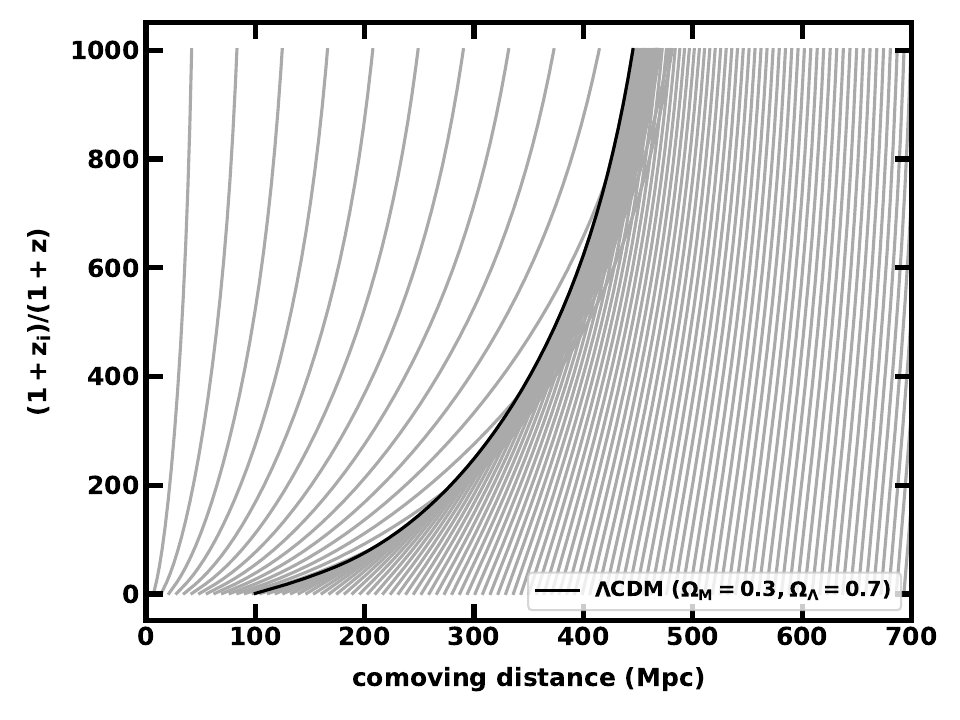}
    \caption{Mass flow lines in a uniform low-density region as a function of the scale factor $a$ for a $\Lambda$CDM cosmological model with starting conditions similar to reference~\cite{blumenthal1992}. The density inside the region is 10\%\ less than that of the surrounding material at $z=1000$ and all mass elements initially follow the Hubble flow. Each gray line tracks the path of a thin spherical shell. The shells inside the low-density region expand more rapidly than those outside and quickly overtake them. The black curve shows the boundary between the emerging void and the background model where a high-density zone builds and persists.}
    \label{fig:voidexpand}
\end{figure}

Figure~\ref{fig:voidexpand} illustrates of the paths of shells in an extremely deep void forming within a \LCDM\ model. 
\myadd{The example tracks shells in an underdense region starting at a redshift of $z_i = 1000$ with an outer comoving radius $x_v = 100$~Mpc, a constant $\epsilon = 0.1$, and peculiar velocity here set to $\dot{x} = 0$ everywhere.}
\myadd{Although the evolution of a spherical tophat void in this Newtonian approximation is scalable to any radius, we choose these specific starting conditions} to facilitate comparison with figure~1 in reference~\cite{blumenthal1992}, showing calculations with similar initial conditions. Although this underdense region is unrealistically deep for voids of any scale, much less for one of this large size, its evolution poses a clear, challenging nonlinear problem as a numerical benchmark.

For a spherically symmetric \ideal\ void, the ``shell-crossing radius'' --- the radius of the innermost shell that has crossed with another larger shell --- is a natural definition of the void radius; prior to shell-crossing, we use the radius of the shell initially at the outer boundary of the 
underdense region. The dark solid curve in figure~\ref{fig:voidexpand} illustrates how the void radius evolves using this definition, providing an accurate picture of the void evolution that includes nonlinear dynamics and flow of material from the void as it mixes with the background universe.

\subsection{Comparing approaches to \ideal\ void evolution}\label{sec:compare}

With an accurate numerical method for tracking the nonlinear evolution of an \ideal\ void, we can assess how well the linear theory and the Zel'dovich approximation perform. For this assessment, we consider a void with depth profile
\begin{equation}
    \epsilon(r) = \left(1 - \frac{2r^2}{3r_v}\right) \exp^{-r^2/r_v^2}, 
\end{equation}
where $r_v$ characterizes the void's size. This profile corresponds to a compensated void, meaning that the relative underdensity in the void is balanced by an overdensity just outside of it, so that the mean mass beyond a few times $r_v$ is equal to that of the background universe. This choice simplifies the use of fast Fourier transforms in our 3-D particle code for the Zel'dovich approximation, foreshadowing our numerical method for tracking more realistic voids. 

\begin{figure}[htbp]
\centering
\includegraphics[width=6in]{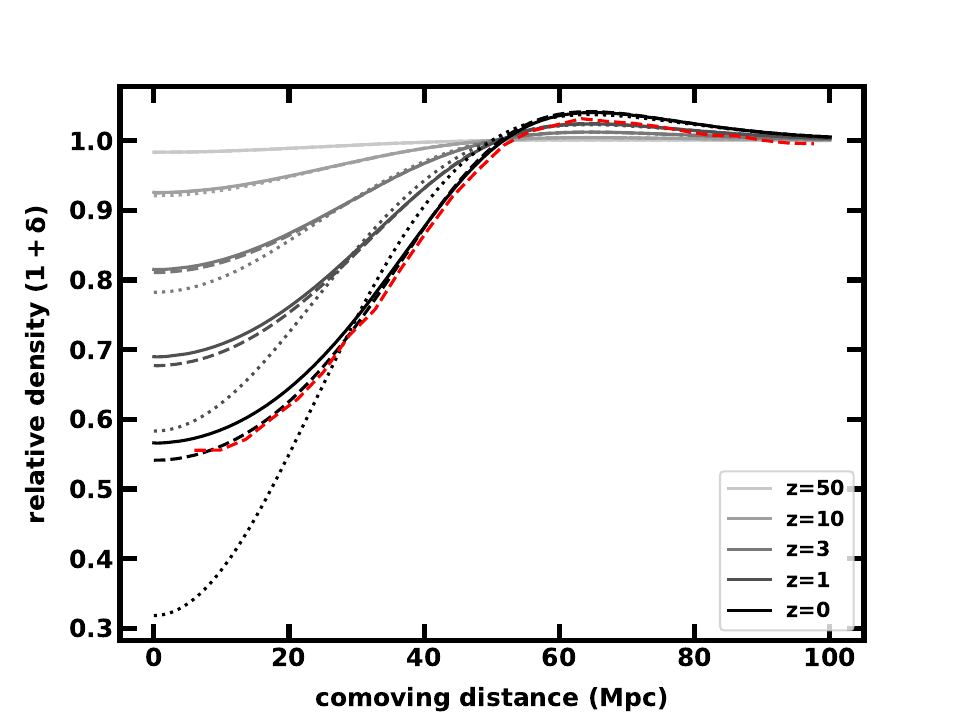}
    \caption{
    Relative density in a Gaussian-compensated spherical void. The solid curves show the relative density profiles ($\rho(r)/\overline{\rho}$) from direct integration at various redshifts (section~\ref{sec:shell}). We include the linear theory predictions (dotted lines; section~\ref{sec:lintheory}) and predictions from the Zel'dovich approximation in 1D (black dashed lines; section~\ref{sec:za}) for comparison. The Zel'dovich approximation from a 3D simulation (red dashed line) provides a trace that has modest ``shot noise'' from our use of a grid of points.  Otherwise it  tracks the detailed calculation well.}\label{fig:voidcoldspot}
\end{figure}

We calculate the evolution of a compensated void in all three approaches --- linear theory, the Zel'dovich approximation, and nonlinear calculations. Figure~\ref{fig:voidcoldspot} shows results, with snapshots of the density profile at several redshifts. Compared with the nonlinear solution, the linear theory predictions agree also, but only for void depths below $\dep \sim 0.2$. The Zel'dovich approximation yields profiles that agree well with the nonlinear solutions, to within 10\%\ to void depths of $\sim$0.6.

To summarize our comparison of methods, we confirm that linear theory describes void growth to a depth of about $\dep \approx 0.2$. The Zel'dovich approximation, as implemented in grid-based codes, describes void dynamics even when linear perturbation theory breaks down. We rely on this result in section~\ref{sec:cosmoreal}, below. Next, we continue our analysis with the full, nonlinear calculations afforded by the shell model.

\subsection{The evolution of \ideal\ voids}

\begin{figure}[htbp]
    \centering
    \includegraphics[width=6in]{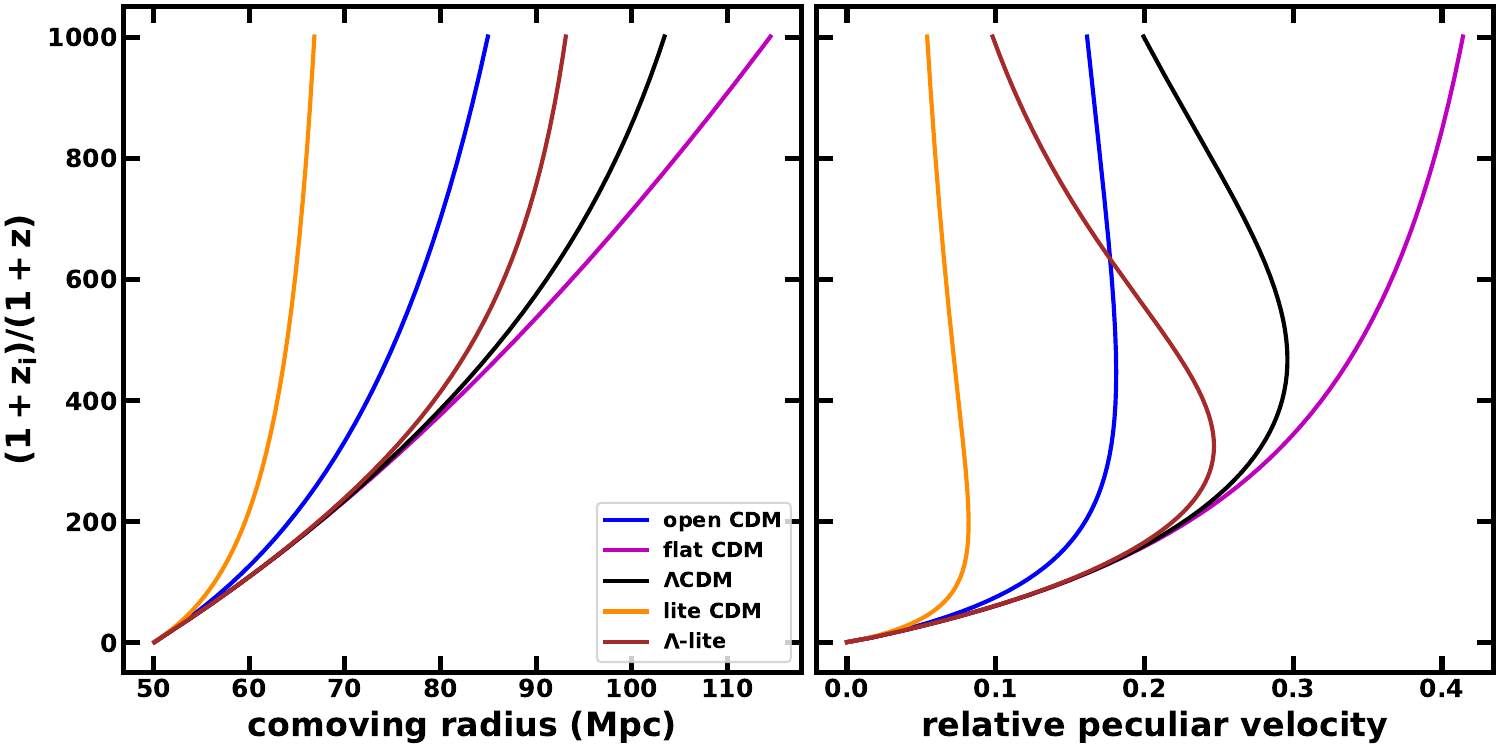}
    \caption{Void sizes and peculiar velocities for constant-density spherical voids expanding into a uniform background with the mean cosmic density. The void edges in the left panel correspond to  simulations that start off at identical redshift ($z=1000$) and relative depth ($\depinit=0.01$), but the background universes differ as indicated in the legend. The right panel shows the peculiar velocity relative to the background expansion speed in the frame centered on the void. Each velocity curve is from a shell starting at the half-radius of its underdense region. For \ideal\ voids, the peculiar motion relative to the Hubble flow is the same for all shells in the interior of the void.}
    \label{fig:voidexpandvpecall}
\end{figure}

With the nonlinear shell model and a range of cosmological parameters, we explore the evolution of \ideal\ voids to illustrate the impact of cosmology on void dynamics. We start a suite of calculations from the same initial conditions (initial redshift and density contrast profile), tracking void evolution with different sets of parameters describing the background cosmology. Figure~\ref{fig:voidexpandvpecall} shows the void radius growth for the cosmological models in table~\ref{tab:cosmic}. It also shows the peculiar velocity --- the radial outflow of matter in excess of the background expansion --- for a point located half way between the void center and its outer edge. In the plot, the velocity is given in terms of the background expansion speed relative to the void center. For ideal voids, this quantity is $\Heff/H - 1$, the same for all points within each void.

The curves in figure \ref{fig:voidexpandvpecall} highlight the following trends:
\begin{itemize}
    \item Voids grow more robustly with higher cosmic mass density; the void in the flat CDM case ($\Omm = 1$) grows to the largest size while the void in lite CDM ($\Omm = 0.1$) is the smallest. 
    
    \item $\Lambda$ augments the void size at the present epoch when other parameters are fixed. The void in the $\Lambda$-lite model ($\Omm = 0.1$, $\Oml = 0.9$) is larger than the one embedded in the lite-CDM model ($\Omm = 0.1$, $\Oml = 0.0$).
    
    \item Void growth relative to the comoving background generally slows with time and eventually stalls altogether. In high-$\Oml$ models, this ``freeze-out'' occurs quickly as the accelerating expansion overtakes the outflow of matter within voids. In high $\Omm$ models, the freeze-out takes a long time.  
\end{itemize}
These trends highlight the relative roles of matter density and dark energy in void dynamics.  It is well-known that voids grow to smaller sizes and  shallower density contrast in low-$\Omm$ models. The open CDM model, once a contender to explain the observed  large-scale structure, fails. The addition of dark energy enhances the growth of voids. As mass shells within a void move outward, they jump into a comoving frame that is expanding faster than without dark energy. This effect increases the proper distance between shells and dilutes their mutual gravity thus driving more rapid expansion. Eventually the background catches up.

\begin{figure}[htbp]
\includegraphics[width=6in]{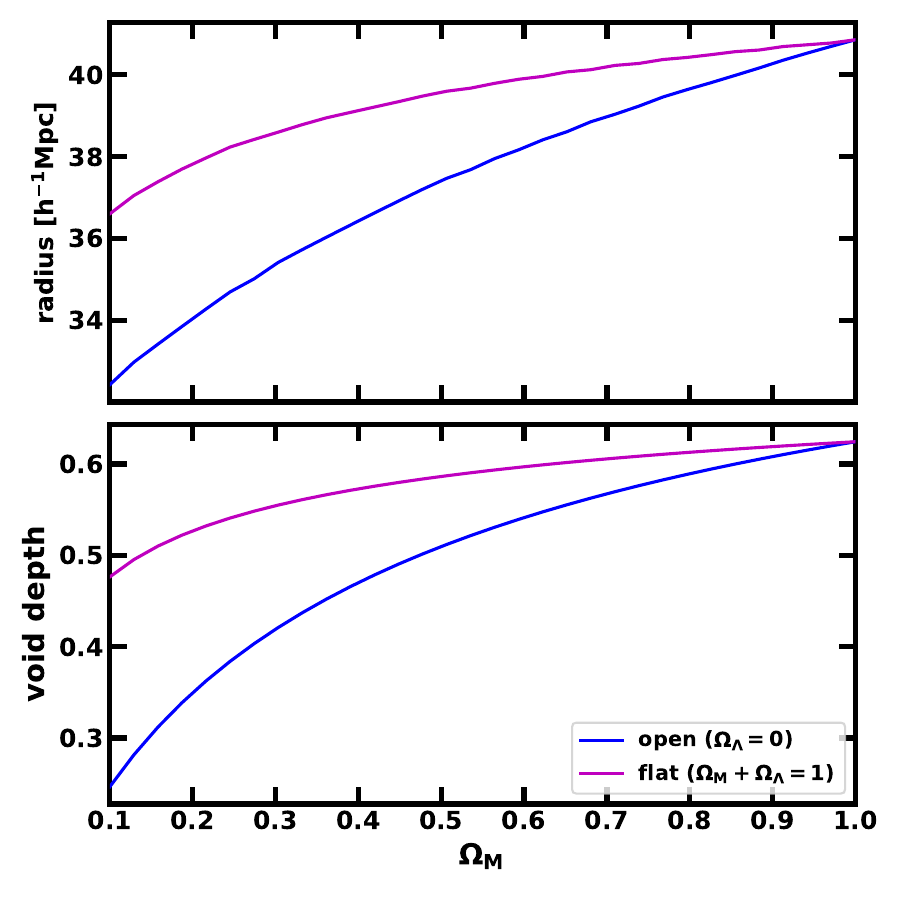}
    \centering
    \caption{Present epoch void radius and depth as a function of $\Omm$ in two sequences of cosmological models. Both the void radii and depth are roughly consistent with the observed power spectrum for initial void sizes of 30~\hinvMpc\ (comoving). Blue curves show open CDM models; magenta curves show geometrically flat models ($\Omm + \Oml = 1$). The void radius increases with increasing $\Omm$. At fixed $\Omm$, models with a positive cosmological constant produce larger voids than for $\Oml = 0$. }
    \label{fig:voidfinal}
\end{figure}

Figure~\ref{fig:voidfinal} gives another view of the impact of both the mass density and dark energy in terms of void depth, $\dep$, in equation`\eqref{eq:dep}, ranging from 0 (no void) to 1 (vacuum). The models shown in curves run from $\Omm = 0$ to $\Omm = 1$, both with and without a cosmological constant ($\Oml = 1-\Omm$ and $\Oml = 0$, respectively). In contrast to earlier plots, this figure tracks voids with sizes and depths comparable to those observed in the cosmic web.

\begin{figure}
    \centering
    \includegraphics{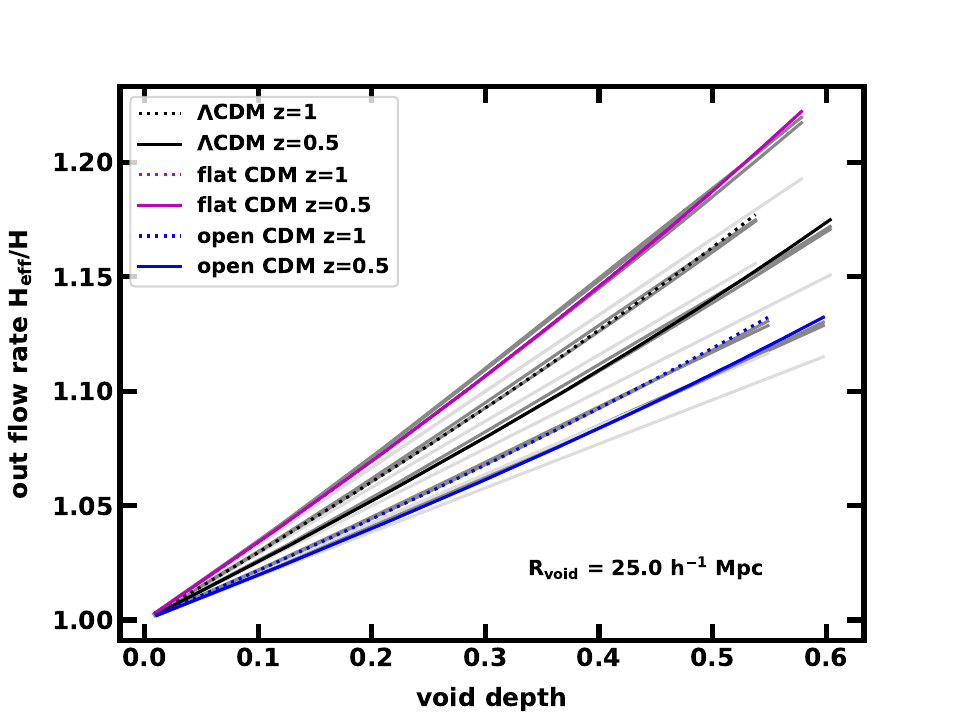}
    \caption{The outflow rate inside a void versus void depth for several cosmological models and at two redshifts. The legend lists the shell simulations; the dark gray curves  show the analytical approximations from equation~\eqref{eq:Heffnonlin} for the outflow rate, formally defined as the Hubble parameter inside a void in units of the background Hubble parameter. Light gray curves are predictions from linear theory.}
    \label{fig:voiddepvel}
\end{figure}

We next examine the dependence of the void depth and outflow rate on cosmological parameters. For \ideal\ voids, both parameters are independent of position within the shell-crossing radius. A hint from figure~\ref{fig:voidfinal} suggests that a combination of void depth (related to void radius by conservation of mass) and outflow rate (derived from peculiar velocities) may contain fingerprints of the background cosmology. Figure~\ref{fig:voiddepvel} provides further evidence. The plots shows the outflow rate as a function of void depth for several cosmological models, sampled at two distinct redshifts. It will turn out that samples of outflow rates and void depths over a range of redshifts are effective indicators of the background cosmology (section~\ref{sec:cozestideal}).

In addition to results from nonlinear calculations, figure~\ref{fig:voiddepvel} shows a simple analytical approximation for the outflow rate as a function of void depth based on linear theory: 
\begin{equation}
   \frac{\Heff}{H} = 1 + \frac{f \dep\left(1+0.22\dep\right)}{3},
   \label{eq:Heffnonlin}
\end{equation}
where the void depth $\dep$ is the measured value. This expression, derived from fitting to a subset of cosmological models \myadd{in table~\ref{tab:cosmic}}, extends equation \eqref{eq:Hefflin} to voids with depths up to $\sim 0.5$, accurate to within 0.3\%\ of the nonlinear results in all the cosmological models \myadd{listed in the table}. 
Coupled with accurate approximations to the growth parameters \cite{lahav1991},
\begin{gather}    
    D  \approx 
        \frac{\Ommz}{(1+z)\Omm} \ \cdot \  
        \frac{\Omm^{4/7}-\Oml+(1+\Omm/2)(1+\Oml/70)}{\Ommz^{4/7}-\Omlz+[1+\Ommz/2][1+\Omlz/70]}, 
        \\
   f  \approx  \left[\frac{(1+z)^3 \Omm H_0^2}{H(z)^2}\right]^{0.6},
\end{gather}
the approximation in equation \eqref{eq:Heffnonlin} leads to computationally fast predictions of void properties from cosmological parameters. This feature is important to enable the dense sampling required by model parameter fits that we use extensively below. 

We also present an analytical approximation for the void depth $\dep$, determined from nonlinear calculations, as a function of the linear theory prediction, $\deplin$ (equation~\eqref{eq:deplin}):
\begin{equation}\label{eq:depnonlin}
    \dep  =  \deplin \left[1-0.60\deplin \left(1-0.27\deplin\right)\right].
\end{equation}
This expression, like equation~\eqref{eq:Heffnonlin}, was also derived from fitting with a subset of cosmological models. It is accurate to within 3\%\ for $\dep \lesssim 0.5$ in all the cosmological models considered here. (See reference \cite{bernardeau1994} for an elegant derivation of a similar relationship.) 

\myaddbegin
As with void outflow rate and depth, linear theory does not accurately predict void radius. In linear theory, a void deepens as mass within it streams through its outer boundary, but the void's comoving radius, $R_\text{co}$, does not change; its size is tied to the Hubble flow. In another view, a void acts as an isolated mini-universe. The void's precursor, with depth $\dep_\text{init}$ and comoving radius $R_\text{init}$ at some high redshift $z_\text{init}$, expands more rapidly than the Hubble flow, increasing in depth as a result. A more realistic scenario involves both mass loss and radial expansion. To include both effects, we use the precursor's depth to predict $\deplin$, the void depth from linear theory (equation~\eqref{eq:deltalin}) to find an upper bound to the mass {lost} from the void.
Assuming that the difference between $\deplin$ and the true void depth $\dep$ stems from the additional expansion of the void, we get a lower bound to $R_\text{co}$. Combining these results, 
%
%
\begin{equation}
\left(\frac{\deplin}{\dep}\right)^{1/3} 
\lesssim
    \frac{R_\text{co}}{R_\text{init}} 
\lesssim
\left( \frac{1}{1-\dep}\right)^{1/3},
 \end{equation}
where the lower limit 
is our estimate with mass loss from the void interior, while the upper limit is from the isolated void model and mass conservation. Our numerical experiments with the shell model confirm that void radii lie between these two bounds.
\myaddend


\subsection{The largest voids}

A potentially powerful constraint for cosmological models is the presence of large voids.  Here, we explore what to expect in various cosmological models, focusing on estimates of the void depth as a function of radius. For these estimates, we use the power spectrum to identify characteristic void depths, and consider the evolution of \ideal\ voids in the shell model. At some large scale, $\lamlin$, we anticipate that $\dep\lesssim 0.2$ for all voids, shallow enough so that linear theory applies (cf.~figure~\ref{fig:voidcoldspot}). We next estimate that length scale.

To determine the linearity scale $\lamlin$, we measure the statistic $\sigR$, the typical relative density fluctuation amplitude in a tophat window function of radius $R$, given the power spectrum of a cosmological model (equation~\eqref{eq:sigR}). The models we consider are normalized so that the linearity scale is somewhat larger than 8~\hinvMpc\ at a redshift $z=0$; for ``1-$\sigma$'' fluctuations in a \LCDM\ universe, $\sigR$ dips below 0.2 at a length of just under 40~\hinvMpc. However, voids this large may arise from more rare, deeper fluctuations. In a volume of radius $R_\text{3G} \equiv 3,000$~\hinvMpc, there are approximately $R_\text{3G}^3/4 R^3 \sim 10^4$ non-overlapping samples of the density field convolved with a tophat of radius $R = 40$~\hinvMpc. Assuming Gaussian statistics and statistical independence of samples, we expect a ``one-of-a-kind'' fluctuation of $3.7\times 0.2$, or, equivalently, a void depth of over 0.7, well beyond the reach of linear theory (figure~\ref{fig:voidcoldspot}). Thus $\lamlin$ must be considerably larger.

\begin{figure}
    \centering
    \includegraphics{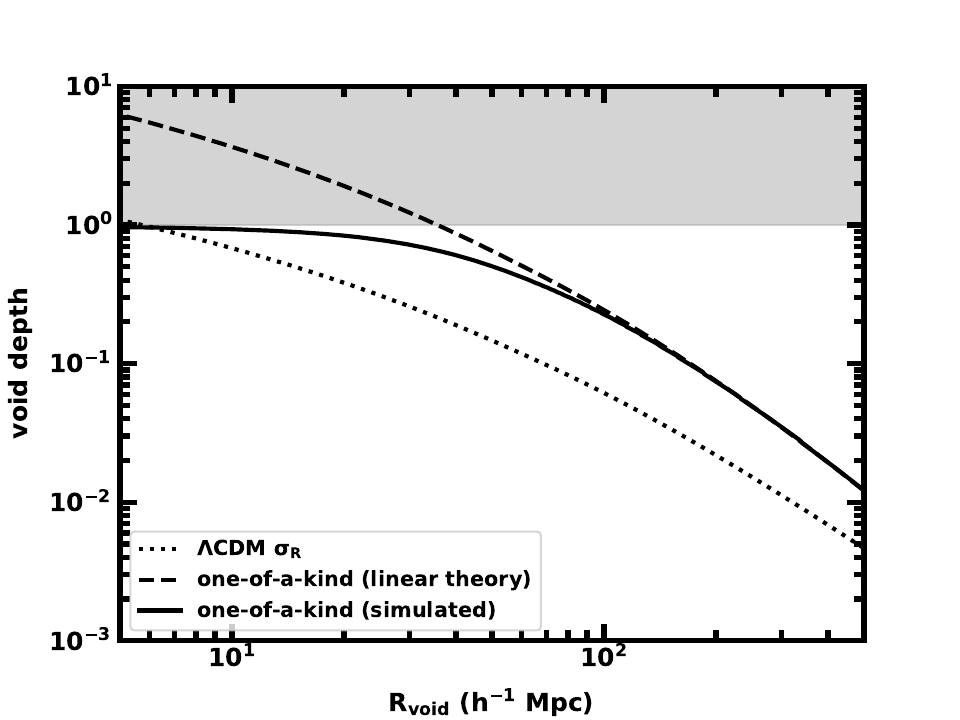}
    \caption{Void depth as a function of void radius for spherical tophat underdensities at redshift $z=0$ in the \LCDM\ model. The $\sigR$ curve (dotted line) shows the linear theory prediction of void depth for a 1-$\sigma$ fluctuation. The ``one-of-a-kind'' curves are the prediction of void depth for fluctuation amplitudes that are rare. We expect only one such void in a survey volume with radius 3,000~\hinvMpc. Linear theory noticeably deviates from the shell simulation results when the void depth exceeds 0.2. The breakdown occurs at a radius around 90~\hinvMpc.}
    \label{fig:sizedepth}
\end{figure}

Figure~\ref{fig:sizedepth} provides a comparison of void depths from 1-$\sigma$ fluctuations at $z=0$, one-of-a-kind void depth from linear theory, and the one-of-a-kind depth when evolved numerically with the shell simulator. As in figure~\ref{fig:voidcoldspot}, linear theory apparently breaks down at a void depth $>0.2$. For the \LCDM\ normalization  we adopt, a good choice for $\lamlin$ is $\sim$90~\hinvMpc. This length is similar in the open CDM model and smaller, about 70~\hinvMpc\ in the other models we explore as a result of the choice of (low) power normalization at translinear length scales.

This analysis suggests that a search for the largest voids in the universe  should rely on linear theory only for scales larger than $\sim$90~\hinvMpc. At smaller scales, the predictions of direct simulations provide more robust comparisons between theory and observations. 

\subsection{Cues from \ideal\ voids}

Simulations of \ideal\ cosmic voids based on the shell model and equation~\eqref{eq:ddotr} provide a testing ground for understanding voids and their connection to the background cosmology. The main results are:
\begin{itemize}
\item  The evolution of void size, depth and radial outflow reflects the background cosmology. The detailed evolution of these properties is well approximated by linear theory for shallow voids (depth $\dep \equiv -\delta \lesssim 0.2$); the Zel'dovich approximation gives more accurate results for deeper voids.
\item The depth of the largest voids, related directly to the power spectrum of fluctuations, is well approximated by linear perturbation theory only for scales above a threshold, $\lamlin$, around 70--90~\hinvMpc.
\item Void depth and outflow rate are simply connected by an analytical expression, equation~\eqref{eq:Heffnonlin}, a modest extension to linear theory. This expression allows estimates of $f H$ (the velocity growth parameter times the Hubble parameter of the background universe) as a function of redshift, independent of void radius.
\end{itemize}

The analytical expression in equation~\eqref{eq:Heffnonlin} is a central result. In principle, with samples of $f H$ over a range of redshifts, we can estimate cosmological parameters in a manner similar to studies of Type 1a supernova \cite{kirshner1999}. We next amplify this result.

\section{Cosmological parameter estimation from \ideal\ voids}\label{sec:cozestideal}

The interior regions of \ideal\ voids evolve as isolated mini-universes. They have FLRW parameters of their own. Nonetheless there are two key connections between voids and the background cosmology. The first connection is the void depth $\dep$, which relates the matter density inside a void to the mean density of the universe. The second connection  is equation~\eqref{eq:Heffnonlin}, which links $\dep$ and the outflow rate $\Heff$ to $f H$, where $f$ is the velocity growth parameter (equation~\eqref{eq:f}) and $H$ is the background Hubble parameter. Just as samples of $H(z)$ from Type Ia supernovae yield $\Omm$, $\Oml$ and $h$,  samples of $f H$ from void properties $\dep$ and $\Heff$ can provide cosmological constraints. 

Here we infer the parameters of the background cosmology from properties of \ideal\ voids. Our approach is based on sets of numerically evolved \ideal\ voids with depths and outflow rates measured at multiple redshifts. We compare these measurements and the predictions of cosmological models from the approximation for void growth in equation~\eqref{eq:Heffnonlin}. To quantify this comparison, we adopt a likelihood function, based on the sum of square residuals between measured void properties and model predictions. We use a maximum likelihood algorithm to find best-fit parameters of the background cosmology. Because the approximation in equation~\eqref{eq:Heffnonlin} is not exact, we include a low level of artificial noise in the analysis so that this small difference does not affect our interpretations. Maps of the likelihood function in parameter space reveal the general sensitivity of parameter estimation to sampling strategies that include the choice of the number of voids and the redshifts when they are sampled. This approach allows a demonstration that simple void properties, when measured for just a few voids over a range of redshifts, provide robust constraints on the background cosmology.

\subsection{Maximum likelihood and MCMC}

To demonstrate that simple void properties lead to estimates of cosmological parameters, for each parameter set we perform $n_v$ numerical simulations with the nonlinear shell model from section~\ref{sec:shell}. We vary the starting conditions and stopping times to obtain a single measurement of $\dep$ and $\Heff$ for each of $n_v$ unique voids. The stopping times sample  redshifts ranging from $z=0$ to $z=2$. To fit the simulated values of $\Heff$ with cosmological model predictions, we define a goodness-of-fit measure
\begin{equation}
    \chi^2 = \sum_{i=1}^{n_v} \left[ y_i^{\text{(measured)}} - y_i^{\text{(theory)}}\right]^2/\sigma_i^2
\end{equation}
where $y_i^{\text{(measured)}}$ are samples of the simulated outflow rate scaled by a factor of 
\begin{equation}\label{eq:Hfid}
    \Hfid = 100 (1+z)^{3/2}~\text{km/s/Mpc},
\end{equation} 
to keep numerical values close to unity. The theoretical values, $y_i^{\text{(theory)}}$\, come from using the measured depths $\dep_i$ in equation~\eqref{eq:Heffnonlin} with a given set of parameters $\Omm$, $\Oml$ and $h$. The ``uncertainties'' $\sigma_i$ in the expression for $\chi^2$ are 
\begin{equation}
    \sigma_i^2 = \sigma_{y,i}^2 + \frac{f^2 \eta^2}{9}\sigma_{\dep,i}^2 ,
\end{equation}
where the $\sigma_y$ and $\sigma_{\dep}$ correspond to noise levels in outflow and depth respectively. The quantities  $f$ and $\eta = \Heff/\Hfid$ are determined for a given model at the redshift of the $i^\text{th}$ void. The noise model adopted here is that measurements have  normally distributed fractional errors at a level $\relerr \sim 0.001$--0.1. By including the noise at some tunable level, we can identify correlations in fitted parameters that are clearly the result of the cosmology and not numerical artifacts or the result of our use of an approximation for theoretical predictions. While not our focus here, the addition of tunable noise also allows us to explore the impact of observational errors.

We use a maximum-likelihood approach to fit measurements with model parameters. In this framework, a likelihood function $\cal{P}$ returns the probability of obtaining specific measurements ($\dep$, $\Heff$) given a set of model parameters $\Omm$, $\Oml$, and $h$ (the curvature contribution is derived from $1 = \Omm + \Oml + \Omk$). Fitting measurements with models amounts to sampling and maximizing  $\cal{P}$ in parameter space. We use the \texttt{emcee} Markov-chain Monte Carlo (MCMC) package \cite{emcee} to do the sampling, with multiple ``walkers'' mapping out the probability distribution $\cal{P}$ in the domain $\Omm = [0.045,1]$, $\Oml = [0,1]$, $h = [0.5, 0.9]$. We use uninformative priors (all models in this domain are equally likely). We thus distribute the starting location of the walkers in parameter space with a low-discrepancy quasirandom number generator (\texttt{scipy.qmc}). We determine best-fits from the global maximum of the likelihood function as mapped by the MCMC walkers;  we obtain uncertainties and correlations between parameters from the distribution of samples in parameter space.

\begin{figure}
 \includegraphics[width=3.1in]{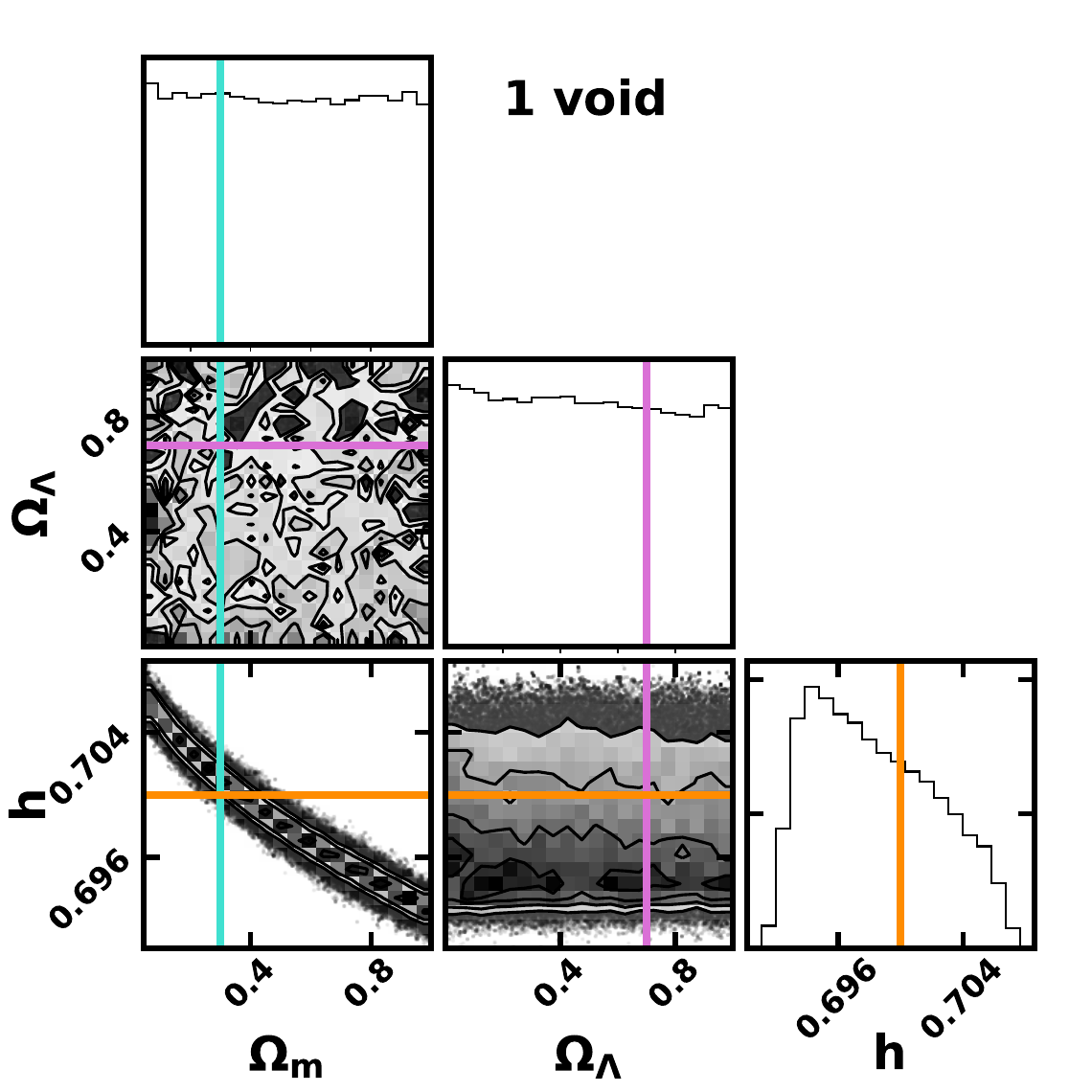} \hfill 
 \vspace{-0.03125in}
 \\
 \includegraphics[width=3.1in]{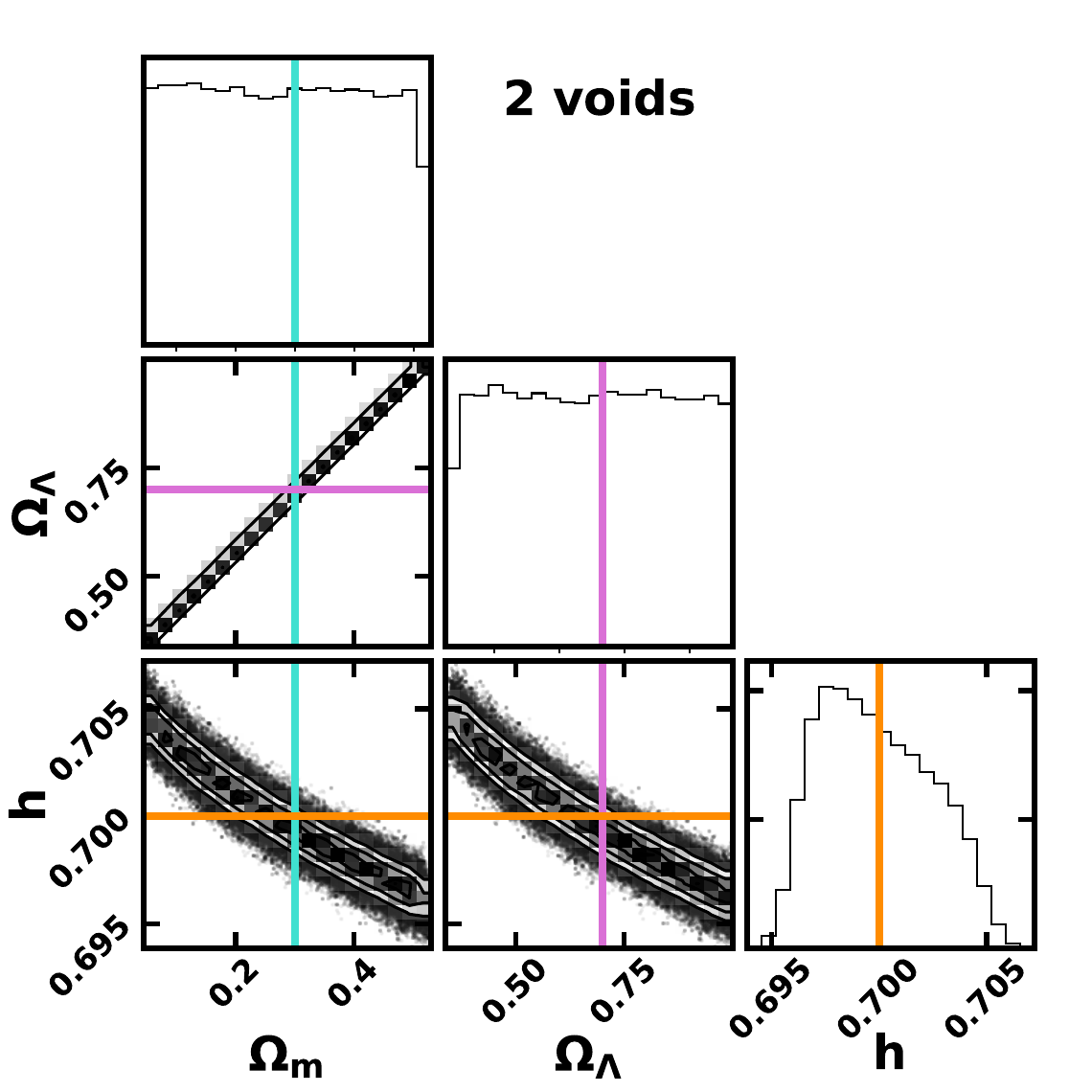}
 \hspace{-0.125in}
 \includegraphics[width=3.1in]{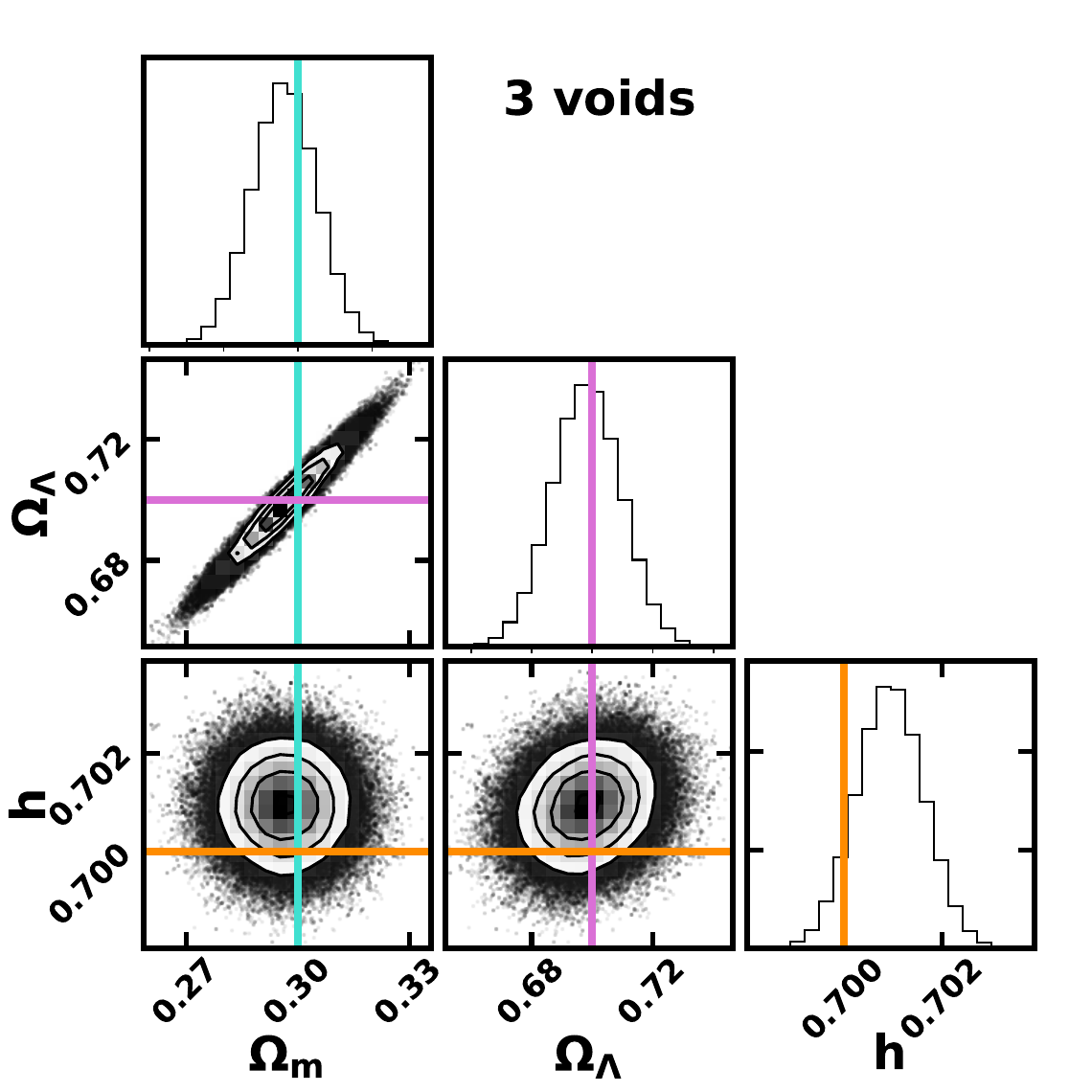}
    \caption{A corner of corner plots \cite{corner} showing probability distributions for parameter fits to void depth and outflow rates. The labels indicate the number of voids measured. The lines locate the true values of the parameters $h$ (orange), $\Omm$ (turquoise), and $\Oml$ (blue).  The fits are most sensitive to $h$, then $\Omm$, and finally $\Oml$. In the upper corner plot, showing likelihood distributions   for a single void, only the Hubble parameter is well constrained. Asserting prior knowledge of $h$ at its true value selects the probability distribution in a slice along the orange lines. With that prior, the (posterior) probability distribution for $\Omm$ becomes sharply peaked around its true value. In the lower left corner plot (two voids, $z=0$ and $z=1$), both $h$ and $\Omm$ are well constrained. In the lower right panel, with three voids ($z=0,0.5,1$), $\Oml$ is also well constrained.
    \label{fig:searchideal}}
\end{figure}

Output from the MCMC sampling highlights the cosmological information contained in voids. The panels showing corner plots in  Figure~\ref{fig:searchideal} illustrate likelihood samples from one, two and three voids, each at a unique redshift. All of the depths and outflows  have small ($\relerr=0.1$\%) uncertainties. With a single void at $z=0$ and one data point $(\dep_1,y_1)$ (upper panel), the likelihood distribution for $h$ is the narrowest compared with the search domain. The void thus contains the most information about the Hubble parameter. The fit is significantly less sensitive to the matter density parameter $\Omm$. However, $\Omm$ and $h$ are strongly anticorrelated. Thus  if $h$ is known \textit{a priori},  $\Omm$ may be more accurately determined. The likelihood distribution for $\Oml$ is flat and there is no apparent correlations with other parameters. We conclude that a single void,  a snapshot in cosmic time, offers little information about cosmic acceleration. 

The likelihood distributions for measurements of two and three voids  (figure~\ref{fig:searchideal}) further emphasize that  voids measured at three or more distinct redshifts are necessary to constrain the cosmological parameters. Two voids  provide weak constraints on $\Oml$, but only with prior knowledge of either $h$ or $\Omm$. 

\subsection{The redshift range}

\begin{figure}
 \includegraphics[width=3.1in]{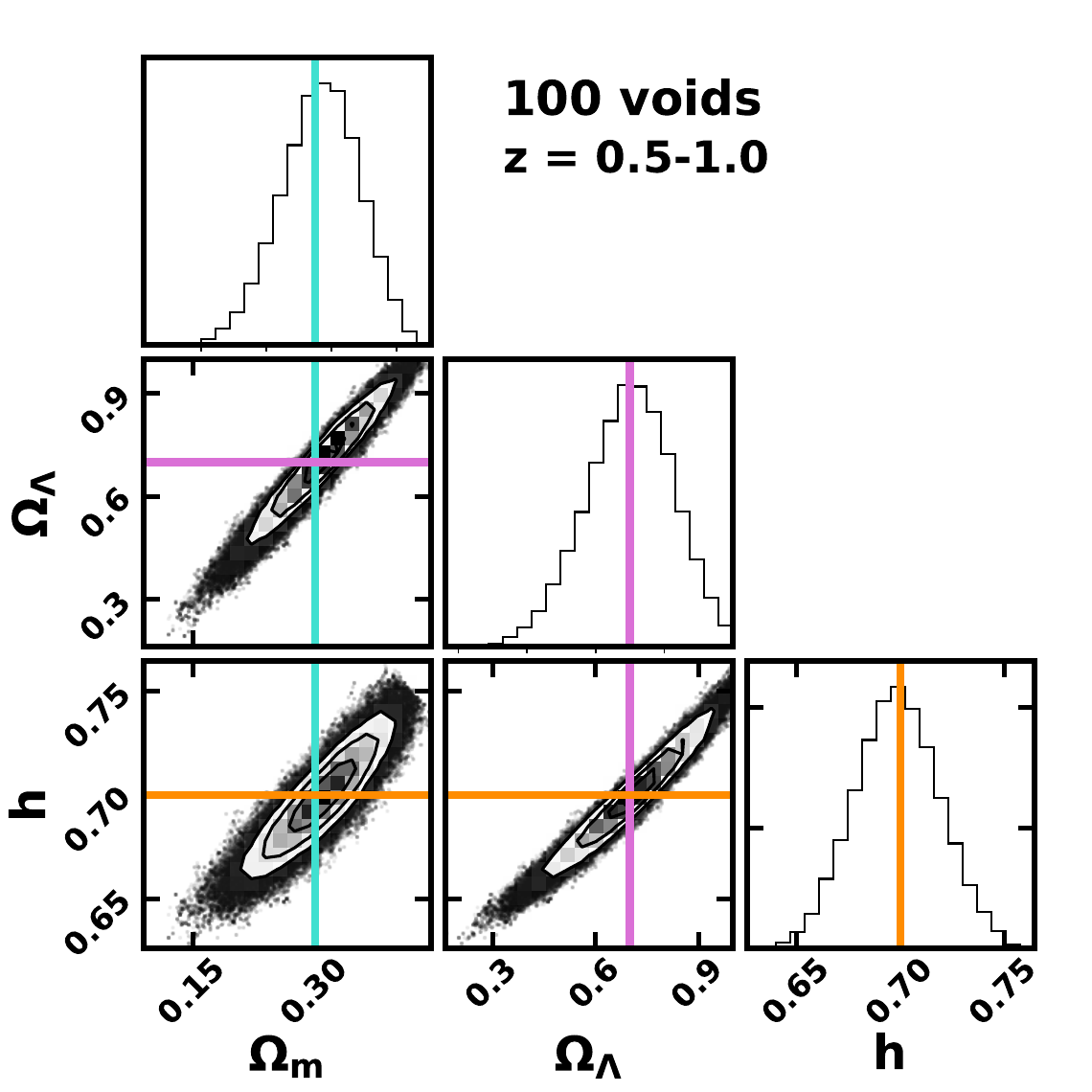} \hfill 
 \vspace{-0.03125in}
 \\
 \includegraphics[width=3.1in]{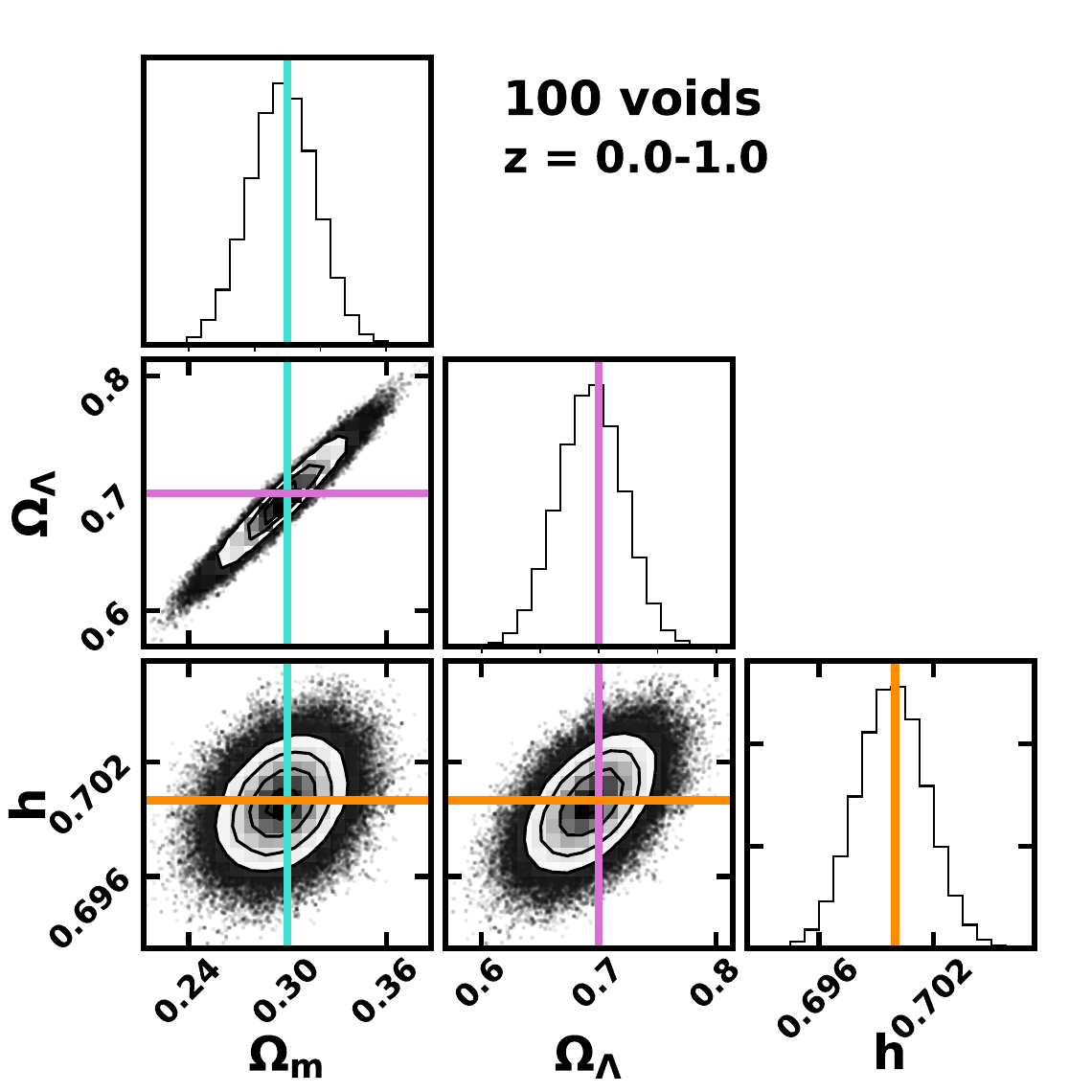}
 \hspace{-0.125in}
 \includegraphics[width=3.1in]{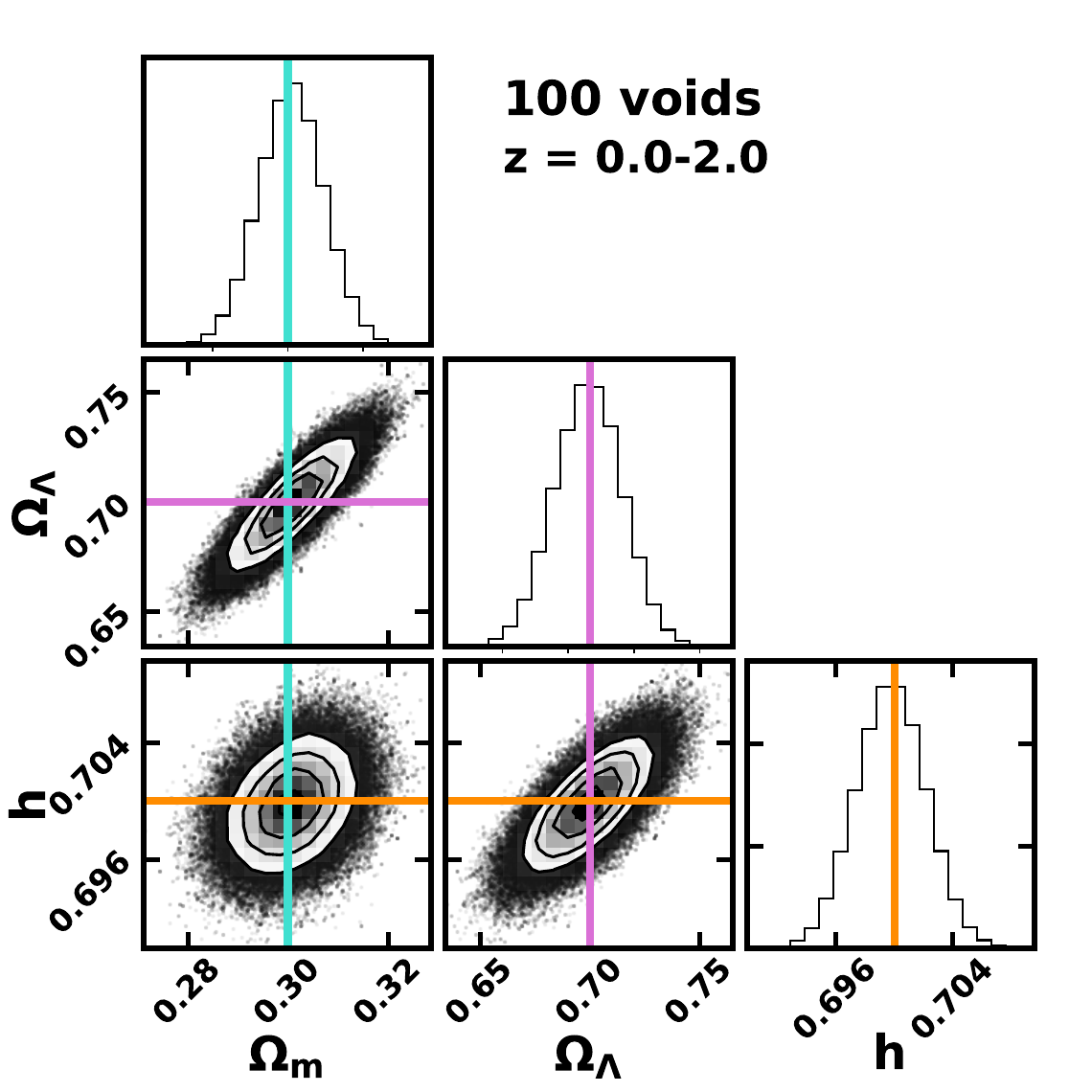}
    \caption{Corner of corner plots with likelihood distributions of cosmological parameters from samples of voids covering different redshift ranges. Each sample includes 100 voids from shell simulations  with artificial noise at the  1\%\ level. From the narrowest redshift range (upper left) to the broadest (lower right), the constraints improve significantly. 
    \label{fig:zdepideal}}
\end{figure}

The Hubble diagram built from Type Ia supernovae provides better constraints on cosmological parameters when applied over a large redshift range \cite{riess1998, perlmutter1999}. 
\myadd{Here, we show that voids similarly provide better constraints when we measure $fH$ as a function of $z$.}
Figure~\ref{fig:zdepideal} illustrates this effect based on samples of 100 voids spaced  uniformly in $z$. The broadest range of redshifts ($z = 0$--2) yields the tightest constraints..

There are two main challenges for measuring voids properties over a wide redshift range. At low redshift, the  observable volume available to sample voids is small; in an absurd limit, there can be at most only one void of radius 100~\hinvMpc\ within a survey of that depth. At high redshift  the challenge is measurement of void properties from faint and distant galaxies. For this demonstration of the theoretical connection between void parameters and the underlying cosmology, we consider only the first challenge, identifying the number of voids we  expect as a function of redshift.

\begin{figure}
    \centering
    \includegraphics{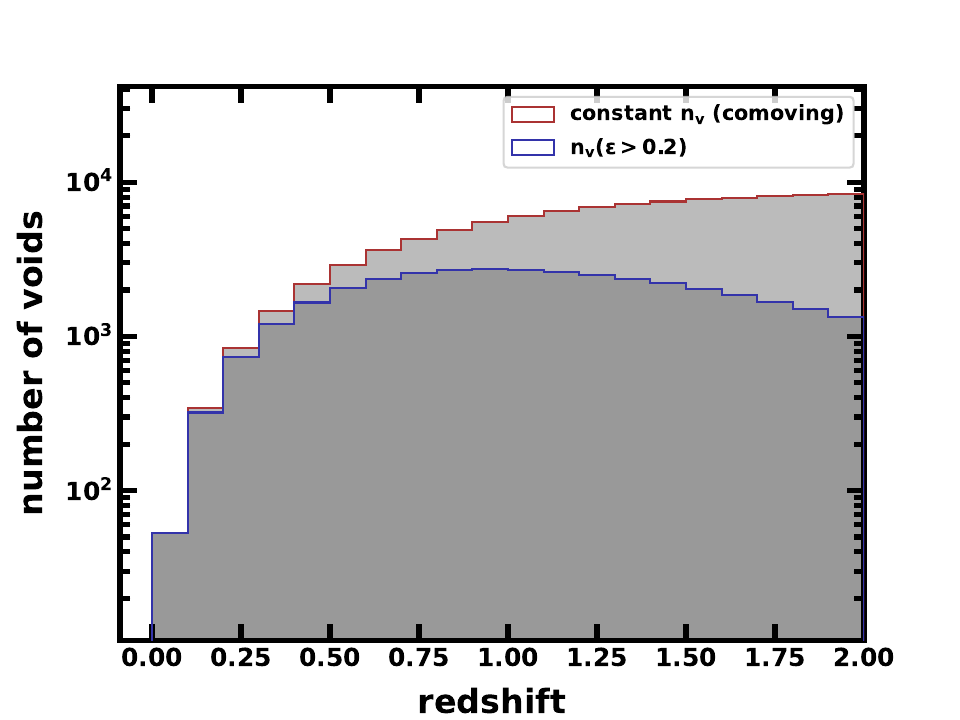}
    \caption{The number of voids in a \LCDM\ universe in redshift bins of width $\Delta z = 0.1$. For this estimate, we use the top 25\%\ deepest underdense regions in the primordial density field smoothed on a scale of 30~\hinvMpc. These regions evolve into present-day voids with a depth of 0.2. The red-colored histogram gives the number of these regions in each redshift bin. The blue-colored histogram shows the subset of these voids that have reached that 0.2 depth at the median redshift of their bin. 
    }
    \label{fig:surveryvolumes}
\end{figure}

\myaddbegin
Our approach for this assessment starts with an analytical estimate of the number of voids within a survey to a limiting redshift of $z=0.1$, a distance of roughly 300~\hinvMpc\ in a \LCDM\ universe. We consider voids with radii of 30--40~\hinvMpc\ and depths $\dep \ge 0.2$, corresponding to regions in the lowest quartile of the mass density distribution smoothed on the void size scale. Although a couple hundred of these voids could fit in the survey volume without overlapping, only about $n_v = 50$ meet the depth criterion. Then, assuming that the comoving number density of voids is constant, we calculate the number of voids that lie in redshift bins of width 0.1 out to $z = 2$. Figure~\ref{fig:surveryvolumes} shows the results; the void counts rise steeply at low redshift ($z \lesssim 0.5$) and increase more slowly at higher $z$, reflecting the 
expansion of the background universe
\cite{hogg1999}. 
We also consider a void census that accounts for void evolution. 
%
At low redshift, 25\%\ of the survey contain voids with $\dep \geq 0.2$ as estimated from the statistics of the smoothed density field; at $z = 2$, this fraction is only a few percent.  
Incorporating this factor when binning voids in figure~\ref{fig:surveryvolumes}, the binned void number counts exceed a thousand for $0.4 \lesssim z \lesssim 2$, with a peak above 2,000 at $z \approx 1$.
\myaddend

\subsection{Summary of cosmology from \ideal\ voids}

By measuring void depth $\dep$ and outflow rate $\Heff$ in voids at multiple redshifts, we  build samples of $f H$ that provide constraints on $h$, $\Omm$ and $\Oml$. This procedure mimics the approach to cosmological constraints based on Type~Ia supernovae \cite{kirshner1999}. Preliminary tests with fits to measurements from solutions to full nonlinear shell model in section~\ref{sec:shell} indeed suggest that samples of void properties over a range of redshifts provide robust cosmological parameter estimates. 

\section{Voids in numerical realizations of the cosmic web}\label{sec:voidsrealish}

The next step in our analysis is to determine how well the approach for \ideal\ voids works in more realistic scenarios. We  use numerical realizations of the cosmic web at large scales and identify voids within them. We then assess the properties of each void in terms of a characteristic depth and outflow rate. Guided by parameter estimation with \ideal\ voids, we use properties from multiple voids covering a range of redshifts to assess the parameters of the background cosmology. The measurements of void properties still have some scatter because they depend on the details of our algorithm for assigning, for example, a single value of depth to a realistic void. This scatter is a source of uncertainty that arises even when we have exact values of the matter density and velocity fields. Future studies will include observational errors in void properties as well, but this analysis is beyond the scope of the investigation here. We limit our discussion to  a demonstration  of the effectiveness of void properties for providing robust constraints on the background cosmology. They have the potential to test of dark energy properties if the background cosmology is already well understood.

We create realizations of large chunks of the universe with an $n$-body particle algorithmm to evaluate the cosmological information encoded by  voids. Our approach is based on  the adhesion approximation \cite{weinberg1990}; similar $n$-body realization methods include the PATCHY code \cite{kitaura2014}.  In our implementation, we use the Zel'dovich approximation in step-wise fashion, allowing  particles to free stream with updates as large-scale structures emerge. Critically, the adhesion approximation adds small-scale viscous diffusion to mitigate the problem of trajectory crossings that arise when free streams of matter in the Zel'dovich flows pass right through one another. As in reference~\cite{weinberg1990}, this feature limits the spatial resolution of the realizations to about a megaparsec. Appendix \ref{appx:adhez} summarizes the implementation.

We adopt the adhesion approximation primarily because it is a fast, accurate way to track void dynamics. Because density fluctuations within voids are low amplitude, nonlinear dynamical effects are weak. Recent work \cite{hamaus2015, stopyra2021, schuster2023} demonstrates that voids are well suited to the Zel'dovich approximation and the related  adhesion algorithm. The realizations of large-scale structure  we construct are geared toward toward modelling voids in the linear and translinear regimes; they are distinct from full $n$-body simulations that track small-scale nonlinear dynamics with Poisson solvers and ODE integrators. 

\begin{figure}
    \hspace*{-0.3in}
    \includegraphics[width=7in]{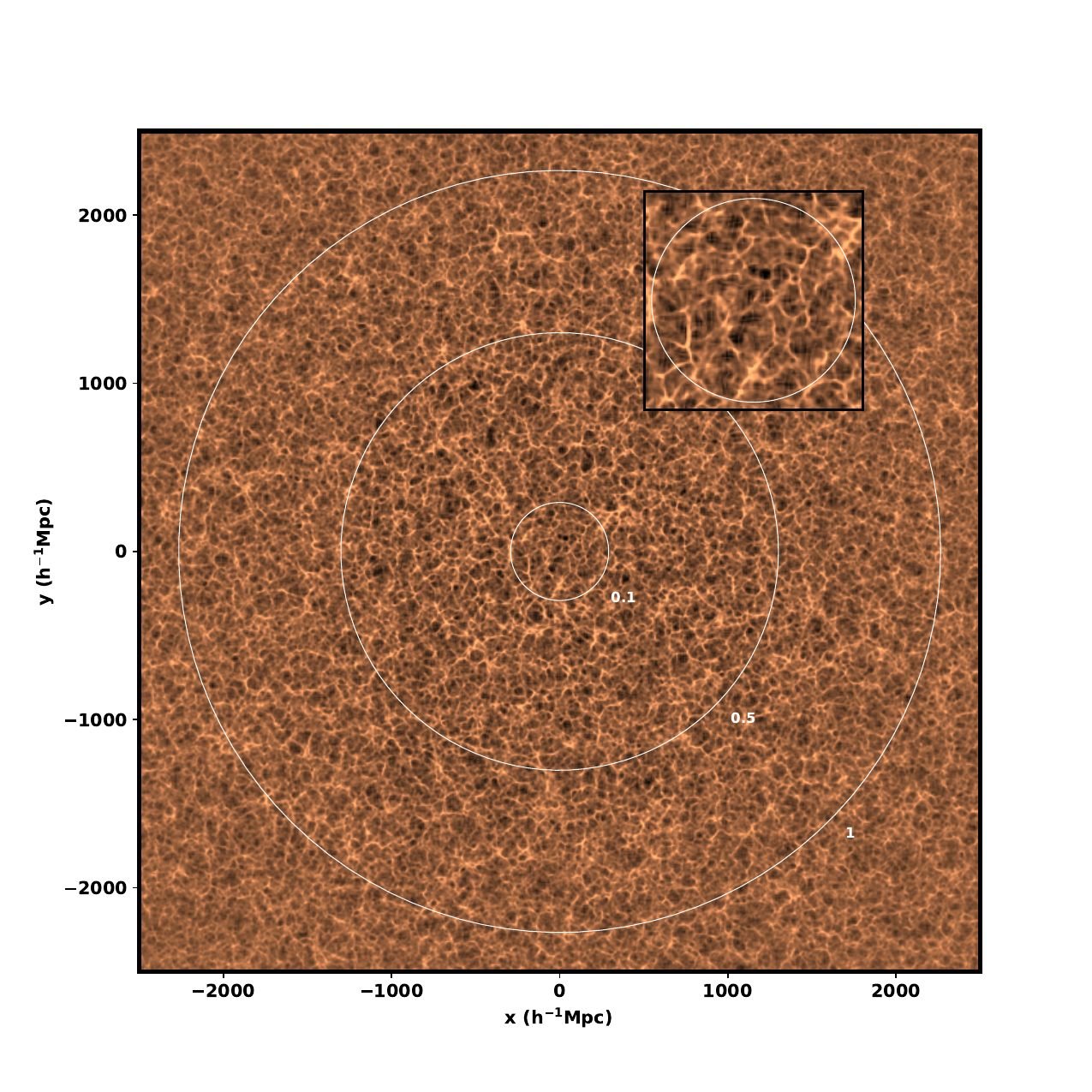}
    \caption{Realization of large-scale structure in a lightcone that extends beyond a redshift of $z = 1$. The color indicates the local mass density in an $x$-$y$ plane, where positions $x$ and $y$ are comoving coordinates with an observer located at the origin. The origin indicates a point in space at the present epoch. Density structures near the origin are evolved to the present time. Moving away from the origin corresponds to moving back in time, or equivalently, increasing redshift, when structures are less evolved. The circles about the origin show the location of points at redshifts 0.1, 0.5 and 1, as labeled. By a redshift of $z=1$, the density fluctuations are noticeably weaker than at the origin. The inset is a zoomed-in view of a region within a redshift of $z=0.1$ (indicated by the circle).}
    \label{fig:adhez5000}
\end{figure}

We generate realizations of the cosmic web for the cosmological models and parameters listed in table~\ref{tab:cosmic} with $n = 256^3$ to 2048$^3$ point particles in cubic regions of 250 to 5000~\hinvMpc\ to a side. Each realization provides a statistically independent snap shot of positions and velocities of mass tracers at a user-specified redshift. In one mode we also construct lightcones where redshift is a function of distance from the center of the realization as in a mock redshift survey. Figure~\ref{fig:adhez5000} shows an example.\footnote{An animation at \href{https://www.astro.utah.edu/~bromley/voidLambdaCDM.gif}{\texttt{www.astro.utah.edu/\~{}bromley/voidLambdaCDM.gif}} shows the growth of large-scale structure in a \LCDM\ universe with the adhesion approximation. We illustrate the emergence of structure in the lightcone in figure~\ref{fig:adhez5000} with an animation at \href{https://www.astro.utah.edu/~bromley/voidlightcone.gif}{\texttt{www.astro.utah.edu/\~{}bromley/voidlightcone.gif}}.} Table~\ref{tab:sims} summarizes the various realizations we performed in this study.

\begin{table}[htbp]
    \centering
\caption{Summary of $n$-body realizations\label{tab:sims}}
\begin{tabular}{lccccl}
\hline\hline
\colhead{cosmology} & \colhead{$L$~\hinvMpc} & \colhead{$n$} & \colhead{trials} & 
\colhead{lightcone?} & \colhead{comment} 
\\
\hline
various & 250-5000 & 256$^3$,512$^3$ & 96 & N & development \\
various & 500 & 512$^3$ & 15 & N & figure~\ref{fig:searchvoidcoz} \\
\LCDM & 2000 & 2048$^3$ & 1 & Y & development \\
\LCDM & 5000 & 1024$^3$ & 3 & Y & development \\
\LCDM & 5000 & 1024$^3$ & 1 & Y & figure~\ref{fig:adhez5000} \\
$\Lambda$-lite & 5000 & 1024$^3$ & 1 & Y & development \\
flat CDM & 5000 & 1024$^3$ & 1 & Y & development \\
\hline
\end{tabular}
\\[1.5pt]\parbox{5in}{\small The models explored here are defined in Table~\ref{tab:cosmic}. The size of the computational grid $n$ equals the number of $n$-bodies in each simulation. The number of trials refers to runs that differ only by the choice of random number seed. The ``Y/N'' flags indicate whether the simulations listed were performed as lightcones (Y for yes) or in a synchronous (constant redshift) frame (N for no, not a lightcone). The comment ``development'' indicates that a realization was performed as part of code development, testing, and cosmological parameter selection; a figure reference in the comment column indicates cases where simulations are incorporated into a figure.}
\end{table}

\subsection{Finding voids}

A key step in assessing the connection between voids and their background cosmology is finding the voids. This process is complex mainly because the cosmic web is not Swiss cheese or even spongelike, with nicely spherical holes embedded in a high-density background. Instead,  choices about density thresholds (voids are not vacuums), sizes and shapes are necessary and lead to a plethora of distinct void-finding algorithms, e.g., \cite{kauffmann1991, elad1997, shesh2004, platen2007, neyrink2008, sutter2015,  cai2015, nadathur2015a, nadathur2015b, shim2023, zaidouni2024}. 

\begin{figure}
    \centering
    \includegraphics{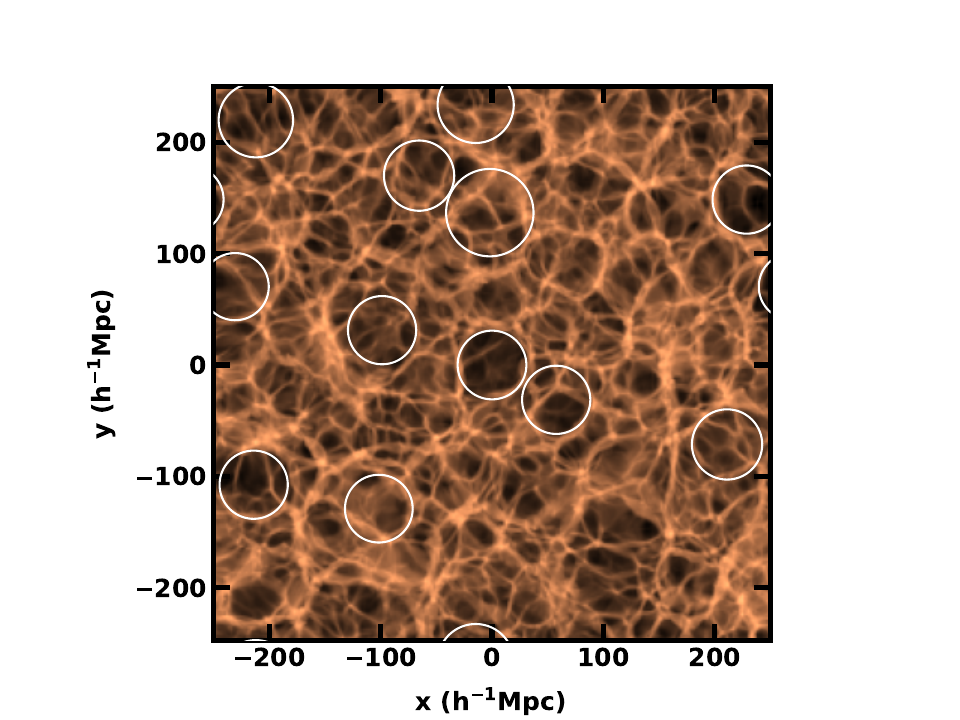}
    \caption{Example of the void finder ($\Rsmoo=30$~\hinvMpc) in a \LCDM\ realization ($L = 500$~\hinvMpc, $n=512^3$). The circles mark voids located within a smoothing length of the slice shown. The realization is shifted to put the deepest void  at the origin.
    \label{fig:voidfinder}}
\end{figure}

\begin{figure}
    \centering
    \includegraphics{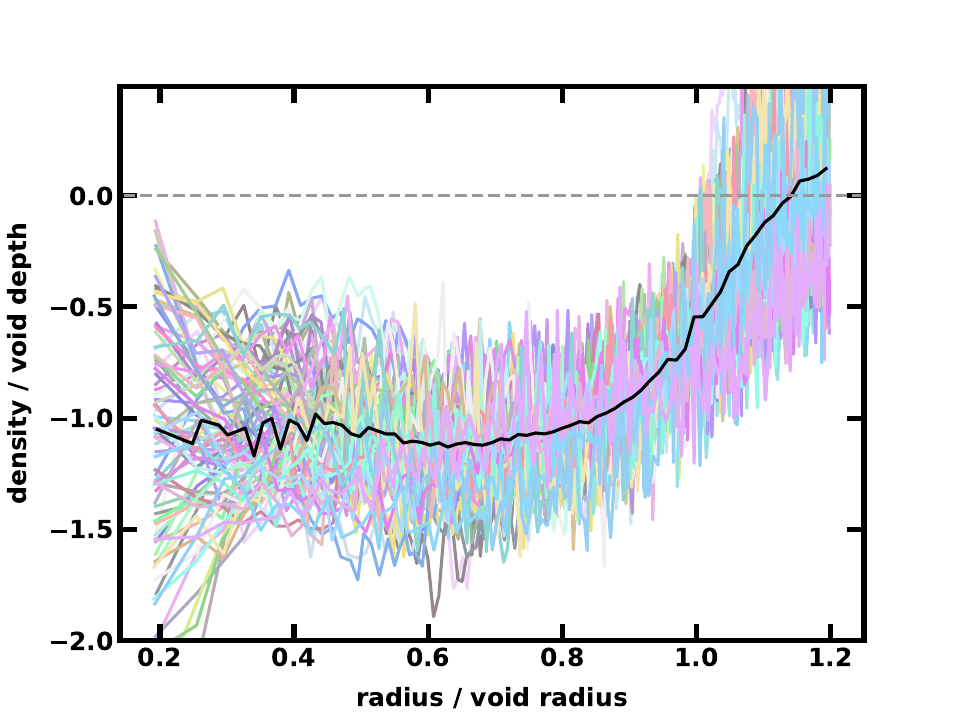}
    \caption{Radial density profiles of voids from the lightcone realization in figure~\ref{fig:adhez5000}. Void radii range 
    from 50~\hinvMpc\ to almost 300~\hinvMpc. 
    Although individual void density profiles (the slurry of light colored lines) vary, on average (black curve) the profiles are similar to tophat window functions with a smoothed step.}
    \label{fig:voidprofile}
\end{figure}

We identify voids in the $n$-body realizations that resemble the spherical shell models in section~\ref{sec:voidsideal} in a straightforward way. Our algorithm specifically targets voids with user-selected radii between $\Rmin$ and $\Rmax$:
\begin{enumerate}
\item \textbf{Interpolate point particles} to estimate the local mass density at points on a regular 3D grid in comoving coordinates of size $L$ per side. With the $n$-body realizations, we use a linear (``cloud-in-cell'') interpolation scheme that respects the periodic boundary conditions of the realizations.
\item \textbf{Smooth the density grid} on a scale $\Rsmoo$ between $\Rmin$ and $\Rmax$. 
For the $n$-body realizations, we use FFT-based convolutions with a 3D tophat window function of radius $\Rsmoo$.
\item \textbf{Identify isolated local minima} in the smoothed density field; each one has a value less than  the mean and is the most extreme in a spherical volume of radius $2\Rsmoo$ centered on it. We use the function \texttt{minimum\_filter} in the \texttt{scipy.ndimage} module for this process.  These minima are void candidates. 
\item \textbf{Apply a mean depth threshold.} To focus on voids of unusual depth, we ignore void candidates with minima less than twice $\sigR$, where $R = \Rsmoo$. 
\item \textbf{Estimate the radius and depth of each void candidate}. This process begins with an estimate of the central void depth,  the average depth within $\Rsmoo/2$ of the location of each isolated minimum. Then we build a smoothed radial density profile by binning particles in concentric shells about the minimum or by averaging adjacent sets of radially-sorted grid values. The void radius, $\Rvoid$, is the distance from the void center where the void profile is half of the central void depth.
\item \textbf{Accept a void candidate} if its radius  is between $\Rmin \leq \Rsmoo$ and $\Rmax$. Otherwise reject the candidate. 
\end{enumerate}
Figures~\ref{fig:voidfinder} and \ref{fig:voidprofile} illustrate the results. Figure~\ref{fig:voidfinder} shows a slice through an $n$-body realization and a few of the  voids that intersect the slice. Figure~\ref{fig:voidprofile} provides radial density profiles of voids calculated from the mass density averaged in concentric spherical shells. Although there is considerable variation from profile to profile, the average radial density profile is characterized by  a flat well with a moderately steep rise near the void edge.

To explore the properties of void depth and outflow rate as in section~\ref{sec:voidsideal}, we select $n$-body particles within each void and measure their number density averaged within a radius $R = \Rvoid/2$. This value is the basis for the depth $\dep$ assignment. The outflow comes from averaging radial velocities of particles relative to the void center in some spherical shell with thickness $0.1 R$. The average $v_r$ divided by $R$  estimates  the outflow rate $\Heff$.

The void-finding algorithm  we use is straightforward and fast. It identifies underdense regions reminiscent of the simple, spherically symmetric \ideal\ voids of section~\ref{sec:cosmoreal} (the void in the center of figure~\ref{fig:voidfinder} is an example).  The algorithm also selects some
voids with complex internal structure or non-spherical shapes. These voids  differ significantly from \ideal\ voids  and produce the ``scatter'' in figure~\ref{fig:voidprofile}. This approach nonetheless connects the idealized spherical void approximations to a more realistic simulation of the observed large-scale structure of the universe.

We choose to implement our own void finding algorithm so that we could most easily assess the impact of changes in the algorithm parameters, balancing code speed, simplicity, and effectiveness at identifying comparatively shallow voids that fit with our analysis. Other options include the adoption of publicly available void finders \cite{villaescusa2018, paz2023}.

\subsection{Cosmological parameter estimation}\label{sec:cosmoreal}

\begin{figure}
    \centering
    \includegraphics{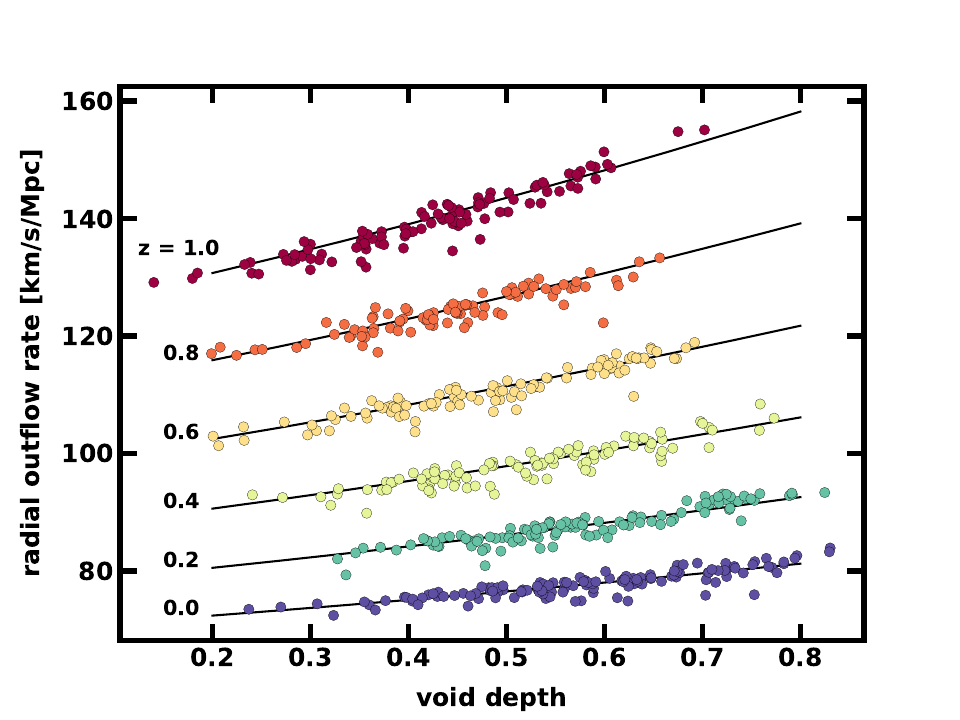}
    \caption{Void depth versus outflow rate for six realizations at various redshifts. The colored points are from $\sim$30~\hinvMpc\ voids identified in $n$-body realizations; the red symbols correspond to redshift $z=1$ and blue for $z=0$, as labelled. The solid curves are the theoretical predictions from linear theory calibrated with the nonlinear spherical shell model from section~\ref{sec:shell}. The theoretical approximations are tuned for shallower voids ($\dep < 0.6$), and their breakdown at higher void depth is evident.}
    \label{fig:voidsummary}
\end{figure}

Following the broad approach of section~\ref{sec:cozestideal}, we measure  two characteristic  void properties, void depth $\dep$  and outflow rate $\Heff$, at the  half radius of each void. With a catalog of voids at various redshifts, we  build a set of samples of these properties. If the void behavior is analogous to  the \ideal\ ones, the sample sets yield estimates of $f H$, where $f$ is the velocity growth factor and $H$ is the redshift dependent Hubble parameter {of the background cosmology}. Figure~\ref{fig:voidsummary} illustrates this approach with $\sim 30\pm 5$~\hinvMpc\ voids in a set of six realizations of the cosmic web at redshifts  from 0 to 1. Although there is variation from point to point compared with the relationships between $\Heff$ and $\dep$ from linear theory, the overall trends are clear.

The theoretical curves in figure~\ref{fig:voidsummary} are fingerprints of the background cosmology. To extract cosmological parameters by fitting the $n$-body samples shown in figure~\ref{fig:voidsummary} with predictions from various cosmological models, we use the same MCMC-based parameter estimators as in section~\ref{sec:cozestideal}: We seek values  in a parameter space of $\Omm$, $\Oml$ and $h$ using a $\chi$-square measure to assess the goodness of fit between theory (equation~\eqref{eq:Heffnonlin}) and samples of $(\dep,\Heff)$ from voids in the $n$-body realizations. Our priors are uniform inside the domain $0 <\Omm < 1.05$, and $-0.1 < \Oml < 1$, expanded slightly from the analysis of idealized voids to better explore the sensitivity of the estimators. The scatter in the samples used  arises from deviations of the $n$-body realization voids from the ideal ones.

\begin{figure}
    \centering 
 \includegraphics[width=3.0in]{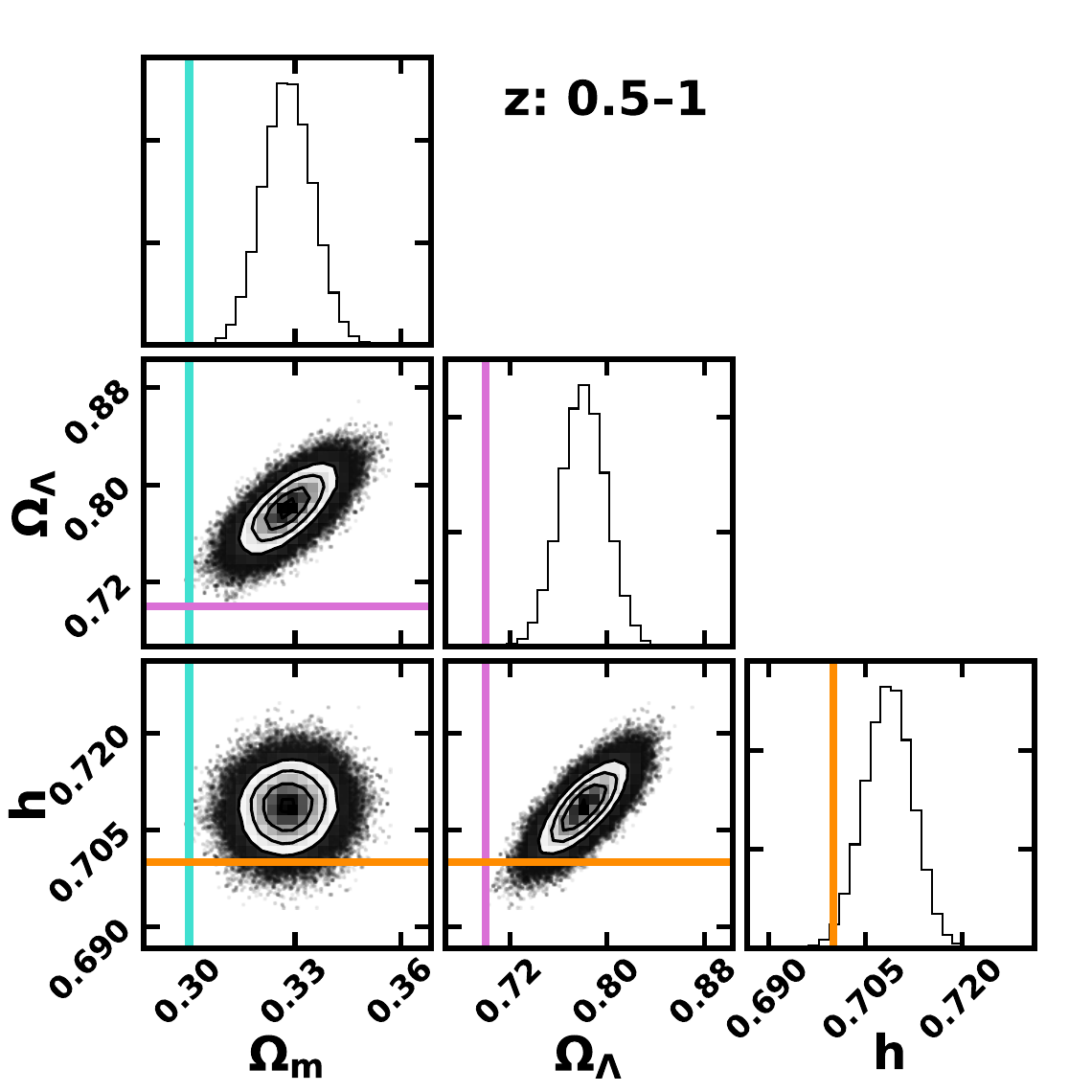} \hfill 
 \vspace{-0.03125in}
 \\
 \includegraphics[width=3.0in]{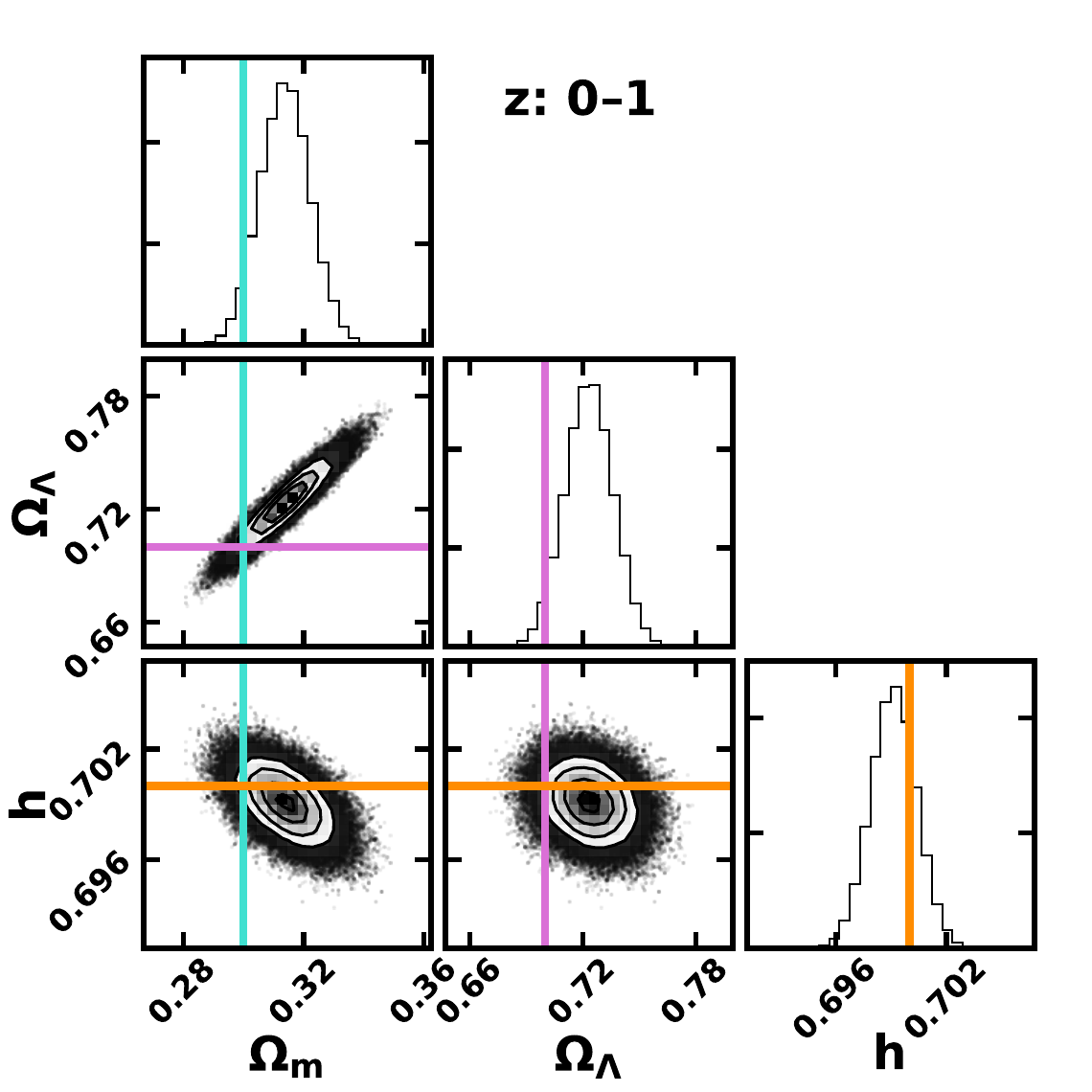}
 \hspace{-0.125in}
 \includegraphics[width=3.0in]{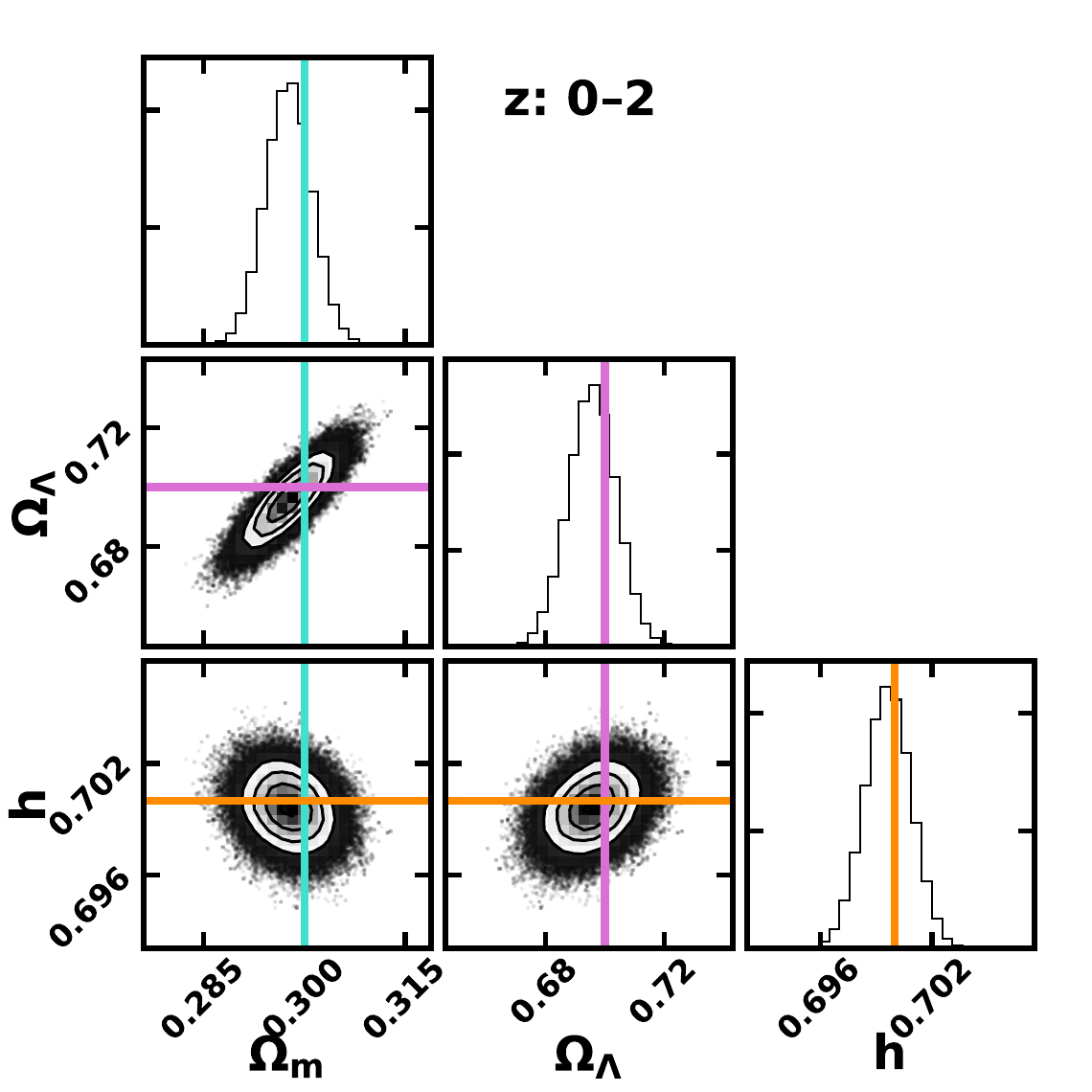}
    \caption{Corner of corners plot \cite{corner} showing MCMC samples in parameter space for independent realizations of the \LCDM\ model at redshifts from $z=0$ to $z=2$. Each realization lies in a cube of 500~\hinvMpc\ to a side and has $n = 512^3$ points. Voids of radii $30\pm 5$~\hinvMpc\ are included. As in figure~\ref{fig:zdepideal} for the  ideal voids, the panels  illustrate that a broader range of redshifts among void samples enhances parameter estimation. \label{fig:searchvoidcoz}}
\end{figure}

Figure~\ref{fig:searchvoidcoz} shows results for the \LCDM\ model (table~\ref{tab:cosmic}). The figure  highlights the central result: void depths and outflow rates are effective as cosmological probes. They offer new, independent constraints on the  parameters that characterize the background universe. The three panels show  different redshift ranges for the samples  and illustrate that the approach is most effective with a broader redshift range.  

\subsection{The largest void in the universe} 

The distribution of void sizes can provide a powerful test of the background cosmology \cite{pan2011}. Here we focus on ``the largest void''. The existence of extremely large voids may place strong constraints on the background model parameters \cite{blumenthal1992}.  The void identification criteria for this investigation are only slightly more stringent than finding local minima in tophat-smoothed density fields \cite{szapudi2015}. Voids in our catalog have profiles that are deep in their central region, and rise to a half-depth value at the nominal radius, behavior that requires specific phase alignment between density fluctuation modes. Supervoids --- composed of multiple smaller voids --- might be missed by our algorithm, depending on their substructure.

Our void finder incorporates a depth threshold filter that requires voids to be at least a 2-$\sigma$ fluctuation in a density field smoothed on a scale comparable to the void size. This cut ensures that voids are significantly underdense. This cut and the requirement that density in a void is consistently low throughout the central regions  favor voids that more closely resemble \ideal\ ones. This aspect is important for applying the formalism of section~\ref{sec:cozestideal}. We  may miss some reasonable voids that could be identified by other algorithms; our search for the largest void errs on the side of caution.

\begin{figure}
    \centering
    \includegraphics{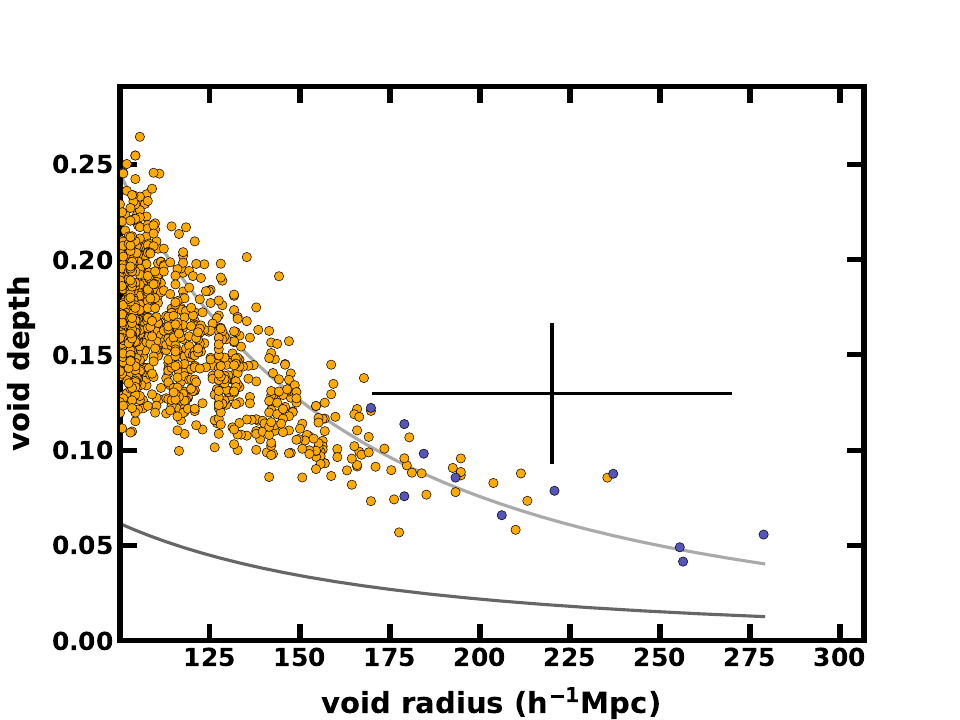}
    \caption{Void depth versus size of the largest voids in a 5000~\hinvMpc\ lightcone realization. This depth-size diagram applies to voids larger than 100~\hinvMpc, scaled according to linear theory. The depth is effectively at the present epoch. The darker solid curve is $\sigR$ from the linear power spectrum; we eliminate void candidates with amplitudes below this value. The lighter-tone line is the amplitude of a one-of-a-kind void predicted in linear theory for the survey volume.  The color of each sample point indicates whether the void is a ``subvoid'' (orange) or a void that is not embedded in a larger void (blue). The error bars show the observed Eridanus supervoid \cite{szapudi2015}.}
    \label{fig:voidsizedistlookback}
\end{figure}

Figure~\ref{fig:voidsizedistlookback} shows  an example of large voids within lightcone realization of 5000~\hinvMpc\ to a side ($n = 1024^3$). Because of grid effects, density modes are poorly sampled for voids with radii $\gtrsim$250~\hinvMpc. All depths are scaled to the present epoch. For comparison,  we also show the Eridanus Supervoid parameters \cite{szapudi2015}. The 1-$\sigma$ errors for the supervoid suggest that it is consistent with voids predicted in the \LCDM\ realization. Since its initial discovery, the supervoid region has been reanalyzed \cite{mackenzie2017, kovacs2022}. The supervoid is estimated to have a greater depth and smaller radial extent, but it is still consistent with standard theory. 

\begin{figure}
    \centering
    \includegraphics{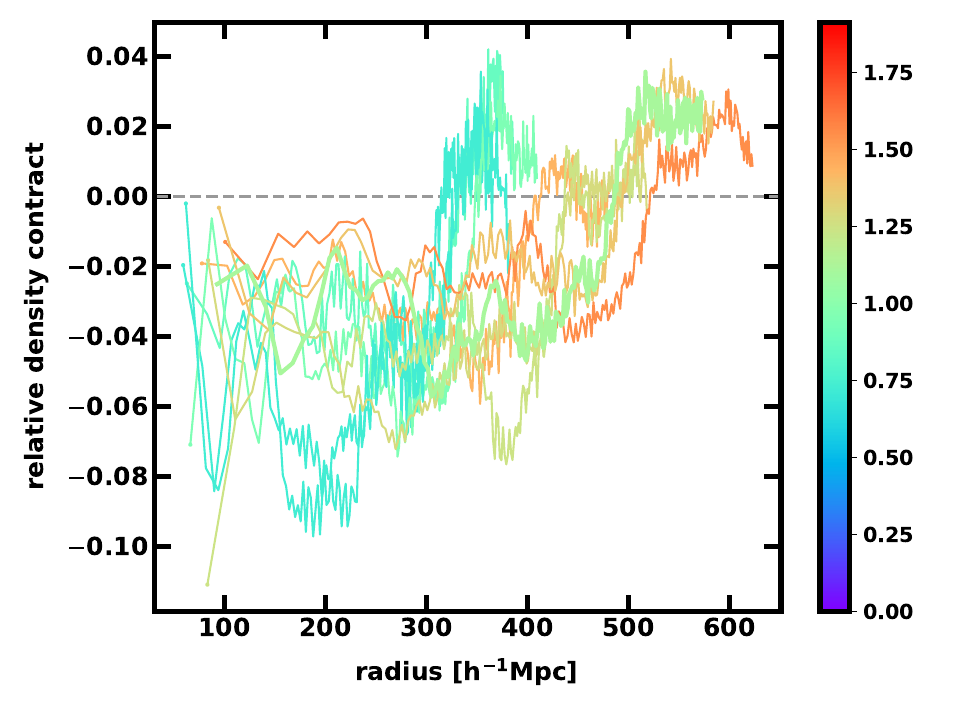}    
    \caption{Radial density profiles of the largest voids in a 5000~\hinvMpc\ lightcone. The panel shows the profile of the top ten largest voids. The color of each trace gives the redshift of each void. The profiles have been scaled to the present epoch.
    }
    \label{fig:adhezprofilelargest}
\end{figure}

In figure~\ref{fig:adhezprofilelargest}, we show the radial density profiles of the largest voids. These profiles are in physical/real space, not  redshift space. 

\section{Voids in redshift space}\label{sec:redshift}

So far, we have used the positions and peculiar velocities of galaxies as tracers to constrain the background cosmological parameters (section~\ref{sec:cosmoreal}). However, observations do not yet measure accurate distances. Instead, extensive galaxy surveys provide the distribution of objects in redshift space. Here we consider the cosmological information content of voids as mapped in redshift space \cite{paz2013, pisani2015, cai2016, nadathur2019zel, nadathur2020prl, correa2021, correa2022, hamaus2022, woodfinden2022, woodfinden2023}. 

An \ideal\ void, with uniform depth $\dep$ and constant outflow $\Heff$, contains galaxies that map from real space to redshift space according to
\begin{eqnarray}
\vec{s} & \approx &  \vec{S} + H\vec{r} + \hat{e}_\text{los} \cdot \vec{v} \hat{e}_\text{los} \\
\ & \approx & \vec{S} + H \left(\vec{r} + \frac{f\dep \Delta_\text{los}}{3}\hat{e}_\text{los}\right)
\end{eqnarray}
where $\vec{S}$ is the redshift-space location of the void center at some distance from the observer, $H$ is the Hubble parameter at that  redshift, and $\Delta_\text{los}$ is a real-space coordinate along the observer's line of sight relative to the void origin. We  assume that the void is observed at a large distance $R \gg \Rvoid$, so that the unit vector along the line of sight, $\hat{e}_\text{los}$, is nearly constant for all points throughout the void. 

The mapping between real space and redshift space significantly impacts the measurement of void properties \cite{kaiser1987}. A small region in real space within an \ideal\ void is stretched in redshift space along the line-of-sight by a factor of $\Heff/H$. The void thus appears deeper in redshift space: 
\begin{equation}
    \deps =  \left(1+\frac{f}{3}\right)\dep 
\end{equation}
This enhancement is substantial --- about 17\%\ for the \LCDM\ model at $z = 0$.  

Void shapes are also affected by redshift space distortions \cite{kaiser1987}. However, the net fractional elongation along the line-of-sight tends to be small because the outflow velocities are small compared with the Hubble flow across the void diameter. In the ideal case, we define the shape of a void in redshift space using the location of particles at the interface with the background universe. Neglecting any shell crossings, we approximate the void shape in redshift space as a prolate ellipse, with ellipticity
\begin{equation}\label{eq:ellip}
    e_f = \Heff/H - 1 \approx \frac{f\dep}{3},
\end{equation}
defined as the difference between the semimajor and semiminor axes, in units of the semimajor axis (also known as the flattening, where $e_f=0$ designates a circle and $e_f=1$ corresponds to a highly elongated ellipse). This result is fragile and depends critically on the spherical symmetry of the underlying real-space density field and on the ability of analysis algorithms to identify the void boundary. 

\begin{figure}
    \centering
    \includegraphics{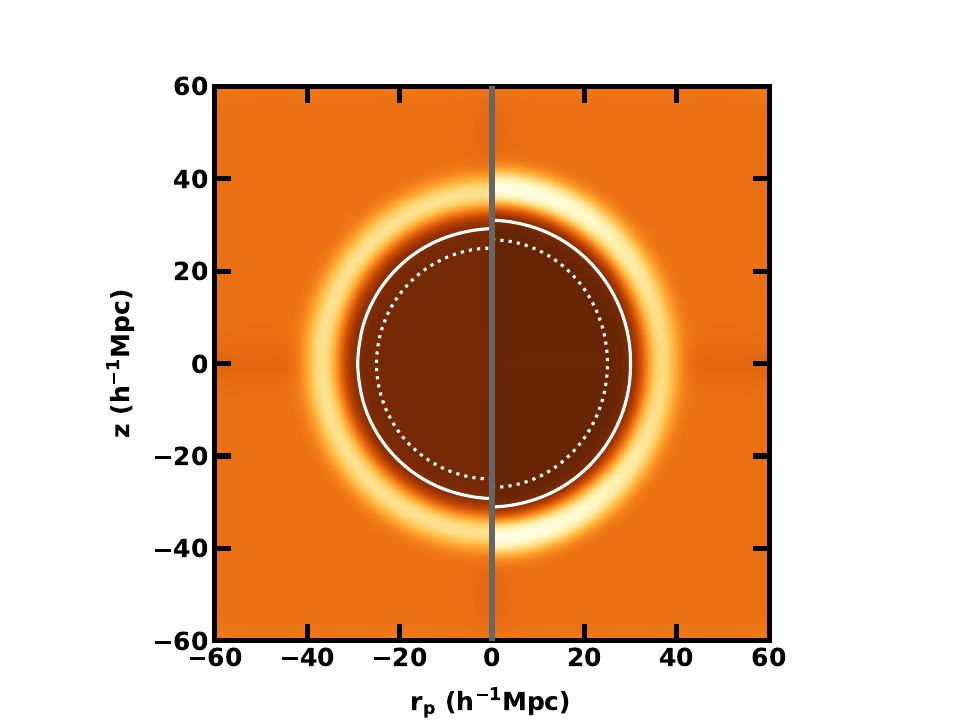}
    \caption{An \ideal\ compensated tophat void in real and redshift space. The real mass density is  on the left; the redshift-space counterpart is on the right. The solid line is an isodensity contour at 50\%\ of the central depth. The dashed line shows positions of points on a circle in real space and its elliptic counterpart in redshift space, assuming that the points comove with the expansion of the void. The image has been smoothed with a Gaussian filter (standard deviation of 2 pixels) to remove grid effects.}
    \label{fig:redshiftideal}
\end{figure}

To highlight the changes in void properties in the transition between real space and redshift space, we consider \ideal\ voids from section~\ref{sec:voidsideal}. Figure~\ref{fig:redshiftideal} shows the real-space and redshift-space densities in the vicinity of an \ideal\ void. The figure offers two lessons: first, the increase in void depth in redshift space from its value in real space is significant. This factor is important in analyses of void properties  inferred from redshift space. The second lesson is that the geometric distortion of real-space patterns in redshift space is, as expected, small. The  void in figure~\ref{fig:redshiftideal}, with a depth in real space of $\dep\approx 0.4$ in a background \LCDM\ cosmology at $z = 0$, has an ellipticity of $\sim$0.07. 

Successful strategies to tease out the small-amplitude distortion of void geometry in redshift space focus on combining data from multiple voids. Aggregate measures like the void-galaxy correlation function quantify averaged redshift-space distortions that emerge away from the void centers, revealing peculiar motion \cite{paz2013}. For \ideal\ voids, this information corresponds to the mean ellipticity at the void boundary. When calibrated with numerical simulations, the void-galaxy correlation function is impressively effective as a cosmological indicator \cite{nadathur2020prl, hamaus2022}. Outcomes include a measure of dark energy with the Alcock-Paczynski test and the velocity growth parameter $f$, as in references \cite{nadathur2020prl, hawken2020, woodfinden2022, wilson2023}. 

\subsection{Reconstruction methods and peculiar velocity surveys}

The large-scale structure revealed in redshift surveys, though affected by line-of-sight peculiar motion of individual galaxies, provides information about the underlying mass distribution in real space. If this real-space distribution were known, then linear theory would predict the peculiar velocities and hence the redshift-space location of galaxies within it. This mapping, at least on scales above $O(10)$~\hinvMpc\ where the theory applies, is invertible. The real-space mass distribution and peculiar velocity flows can be reconstructed from redshift data \cite{yahil1991, dekel1994, fisher1994, nusser1994, gramann1994, tegmark1995, zaroubi1996, taylor1996, hamilton1998, taylor1999, burkey2004, irsic2011, koda2014, davis2014, white2014, watkins2015, carrick2015, howlett2017a, howlett2017b, adams2020, said2020, whitford2022}. Recent work has extended reconstruction methods to include smaller scales where nonlinear dynamics affect redshift-space distortions \cite{hamaus2017, ganashaia2023, qin2023, chen2023, chen2024, lilow2024, wang2024vpec}. 

Density reconstruction methods have been designed for the study of voids \cite{paz2013, nadathur2019zel}. Because the details of void selection are critical in measures of the void-galaxy correlator, reconstruction of the real-space density field prior to void identification is important so that redshift-space distortions do not introduce adverse selection effects \cite{chuang2017, nadathur2019crossx}. 

Reconstruction methods typically depend on assumptions about the background cosmology \cite{kaiser1987, dekel1994} --- the Hubble expansion rate $H$ and velocity growth factor $f$ are often required parameters in a reconstruction application. Therefore, reconstructed density and velocity fields on their own are not well suited for parameter estimation. They return the cosmological information from which they were derived. Following the lead of \cite{dekel1994}, preliminary work suggests cosmological constraints from reconstructed voids at various redshifts are nonetheless possible.  We plan to investigate these issues in future work. 

Independent real-space distance information, combined with redshift data, yield the radial component of peculiar motion. In tandem, these measurement provide meaningful cosmological constraints \cite{qin2023, hollinger2024}. Recent peculiar velocities surveys \cite{tully2016, hong2019, howlett2022} provide estimates for the nearby, low-redshift universe. Upcoming surveys \cite{saulder2023, taylor2023, mould2024, said2024}  offer a deeper view of the cosmic velocity field, including the interior of large voids. 

\section{Conclusion}\label{sec:conc}

Cosmic voids hold a bounty of information about the overall content, structure, and evolution of the universe \cite{alcock1979, kirshner1981, bertschinger1985, white1987, rg91, blumenthal1992, dubinski1993, dekel1994, lee2009, biswas2010}.  Because voids contain comparatively little matter, the effects of dark energy on the dynamics of visible objects lead, in principle, to testable model-dependent predictions \cite{colberg2005, lee2009, tikhonov2009, maeda2011, bos2012}. The mere presence of the ``largest void,''  is one test of models \cite{blumenthal1992}.  

We explore how void dynamics  can constrain the parameters of the underlying cosmology.  We enhance the approaches of \cite{blumenthal1992} and \cite{schuster2023}  where simple models and linear theory are applied to voids. We  also build numerical infrastructure to simulate \ideal\ voids and to generate $n$-body realizations of voids that emerge in  a cosmological context. We focus on voids with sizes larger than 20~\hinvMpc. We use simulation volumes that are between 250~\hinvMpc\ and 5000~\hinvMpc\ to a side. At the largest scale, an individual $n$-body realization provides representative samples of voids with radii up to $\sim$250~\hinvMpc. 

The numerical data here include fifteen independent $512^3$-particle $n$-body realizations (figure~\ref{fig:searchvoidcoz}) and one $n = 1024^3$ lightcone snapshot 
(figure~\ref{fig:adhez5000}). Other data sets in this project include more than 100 realizations with $n = 512^3$, a half-dozen $n = 1024^3$ realizations and one run with 2048 bodies, spanning a range of redshifts, cosmological models (table~\ref{tab:cosmic}) and random number seeds (table~\ref{tab:sims}). The adhesion algorithm (Appendix~\ref{appx:adhez}) is a distinctive feature of this approach because it is at least an order of magnitude faster than full $n$-body simulations \cite{weinberg1990}. The adhesion approximation offers the possibility of maximum-likelihood fitting to void data with realizations evaluated at unique sets of cosmological parameters instead of the analytical approximation in equation~\eqref{eq:Heffnonlin}. Thus, we can include deep voids ($\dep > 0.5$) in parameter estimation even when though the analytical approximation breaks down.

The main results are:

\begin{itemize}
    
    \item \Ideal\ voids demonstrate the dynamical role of dark matter and dark energy in the simple case of a cosmological constant. The trajectories of matter (galaxies) in \ideal\ voids --- the radial position and velocity --- are sensitive to both $\Omm$ and $\Oml$. Each set of parameters produces a unique path in this phase space (figure~\ref{fig:voidexpandvpecall}).
    
    \item We confirm that linear theory and the Zel'dovich approximation provide good descriptions of the dynamics of  ideal voids (figure~\ref{fig:voidcoldspot}).
    Linear theory, with slight modifications, connects void depth $\dep$ and outflow rate $\Heff$ to the background cosmological parameters $H$, $\Omm$, and $\Oml$ through dependence on the overall background expansion and the velocity growth parameter $f$.
    
    \item When measured over a range of redshifts, void properties $\dep$ and $\Heff$ give estimates of the Hubble parameter, $H(z)$, at different cosmological times. As with analyses of Type Ia supernovae, these values in turn reveal the expansion history of the universe as a whole, and thus constrain $\Omm$ and $\Oml$.  We illustrate with both \ideal\ voids and more realistic samples generated with $n$-body realizations based on the adhesion approximation (figures~\ref{fig:searchideal} and \ref{fig:searchvoidcoz}).

\myaddbegin
    \item Our numerical tracking  of \ideal\ void evolution identifies a simple approximation for the nonlinear relationship between the outflow rate $\Heff/H$ and the void depth $\dep$ (equation~\eqref{eq:Heffnonlin}). This computationally efficient and surprisingly accurate formula enables precise  estimation of cosmological model parameters from void properties in shell simulations and in large $n$-body realizations.
\myaddend

    \item  We generate lightcone maps from $N$-body realizations. These lightcones simulate the  large-scale structure observed in redshift surveys. With a simple void finding algorithm, we explore the properties of the largest voids in the  simulated observable universe. Observed ``supervoids'' spanning hundreds of megaparsecs are likely in the \LCDM\ model.

    \item We focus on void properties in real space including  exploration of their physical mass densities and outflows.  This demonstration of principle may eventually lead to observational constraints with much improved direct distance measurements.
    
    \item Current maps of large-scale structure are primarily redshift surveys. We discuss interpretation of the properties of the largest voids in redshift space. Voids appear deeper in redshift space than they are in real physical space (the Kaiser effect; \cite{kaiser1987}).
    
    \item  To forecast future parameter estimation from voids in redshift surveys, we briefly mention adopting real-space reconstruction algorithms that extract physical void depths and outflow rates from redshift survey data \cite{dekel1994, nadathur2019, nadathur2020, nadathur2020prl, aubert2022, woodfinden2022, woodfinden2023, radinovic2023}.  This avenue may eventually be promising at $z \sim 1$ where the number of voids that can be accessed observationally is large. 
\end{itemize}

To explore what we might learn from voids, we study the simulated universe where we have full knowledge of void density profiles and outflow velocities. Simple void properties derived from these quantities constrain parameters of the background cosmology, providing independent tests from low density regions where the dynamical effects of dark energy are enhanced. Current observations can not measure these properties well enough to apply the approach we describe. With improvement of observational capabilities over time, we may eventually realize the full potential of cosmology with voids.

\acknowledgments

We thank an anonymous referee for providing comments that improved the presentation of this work. We are grateful to NASA for time on the Discover supercomputer. The Smithsonian Institution supports the research of MJG.

\appendix

\section{The adhesion approximation}\label{appx:adhez}

The adhesion approximation is an algorithm designed for $n$-body representations of large-scale structure \cite{weinberg1990}. The algorithm's designers provide the details and rationale. We outline the steps to create a single realization with the implementation we use.
\begin{enumerate}
\item Select the cosmological model parameters $\Omm$, $\Oml$, $h$, the linear power spectrum $P(k)$, an initial redshift (e.g., $z_i= 10$--1000), and a final one (e.g., $z_f = 0$). We also specify a unique random number generator seed for each realiation.

\item The adhesion algorithm requires an integration over time, performed using the growth factor $D$ (equation~\eqref{eq:D}) at $z_i$ and $z_f$. We set up a for-loop of ten-ish identical steps of size $\Delta D$ running from $D_i = D(z_i)$ to $D_f = D(z_f)$. To create a survey with redshifts that vary with position (defined on a grid, below), make $D(z_f)$ depend on the real-space grid location. This  step is a new addition to the original algorithm.

\item Define a cubic, triply-periodic computational domain with length $L$ on a side and $n^3$ gridpoints. Because  we  call Fast Fourier Transforms (FFTs), optimal choices for $n$ are powers of two. Because the adhesion approximation is a translinear method, the recommended ( \cite{weinberg1990})  grid spacing is about 1~Mpc.

\item We choose $L = 250$~\hinvMpc\ and $n = 256$ for the smallest production runs. On modern computer hardware without parallelization, this choice generates a single realization in a minute. Runs with $n = 1024$ complete within an hour.

\item Realize a primordial density field in the Fourier domain choosing its normalization at the starting redshift. Define a grid in Fourier space with spacing $dk = n/L$. Calculate wave vector amplitude $|k|$ and evaluate $P(k)$ for each gridpoint. (We select either the CAMB algorithm \cite{CAMB} or the Eisenstein-Hu formulae as in the \texttt{nbodykit} package). The relative density at a gridpoint in $k$-space is
\begin{equation}
\delta(k) = \sqrt(P(k;z)/L^3)\times (u+i v)\times n^3
\end{equation}
where $u,v$ are two independent normal variates (zero mean, unit standard deviation).
The factors of $L$ and $n$ are needed to  scale the amplitude for the Python/Numpy (`\texttt{np}') FFTs correctly. We zero out the $k=0$ mode. 

\item Calculate $\Phi_0(k) = \delta_k/k^2$, related to the gravitational potential, and take its inverse FFT (\texttt{np.fft.ifftn}) keeping only the real part. The density grid is not used again in the prescription and may be deleted.

\item Calculate the three spatial components of $\vec{\nabla}\Phi$ with a centered finite difference; \texttt{np.roll} is useful here.

\item Set up three coordinate grids so that $x,y,z$ are grid positions in the forward (spatial) domain (e.g., x values run from $-L+dL/2$ to $L+dL/2$; \texttt{np.meshgrid} is helpful).

\item Update the grid positions with an initial half-step, $\Delta\vec{x} =  \vec{\nabla}\Phi \Delta D/2$ (equivalent to a Zel'dovich step). 

\item Toward implementing small-scale viscous diffusion, define $u = \Phi_0/2/\nu$ where viscosity $\nu = \alpha dL^2$ and $\alpha = 1$; if $\max(u) > F$, where $F = 702$ for  our double-precision floating representation; choose $\alpha >1$ with the smallest value needed to prevent overflow. Finally, calculate $U_0 = \exp(-u)$. This step impacts the resolution of the smallest scale structures.

\item Implement the integral over $D$: loop over the growth factor from $D = D_i$ to $D_f$, bearing in mind the half-step above. At each step $k=1,2,...$ find the standard deviation $\sigma = \sqrt{2(D_i+k\Delta D)\nu}/dL$ for a Gaussian convolution kernel. If $sigma < 0.25$ continue stepping positions as in a pure Zel'dovich approximation. Otherwise perform the Gaussian convolution of $U_0$ (\texttt{scipy.ndimage.gaussian\_filter} is efficient) and then calculate $\Delta\vec{x}/ \Delta D=2\nu\vec{\nabla}\ln U_0$ using finite-differences. Update the positions accordingly. To complement the initial half-step, the last iteration is a half-step as well.

\item Generate velocities for each displaced grid point at the end of the integration loop, $\vec{v} = Hf\Delta\vec{x}/ \Delta D$.
\end{enumerate}

The results are the components $(x,y,z,v_x,v_y,v_z)$ of displaced gridpoints constituting the phase-space coordinates of individual $n$-body particles. We can reconstruct the density field by interpolating the ``deformed'' positions back onto a regular grid with a linear (or higher-order) interpolator. The novel part of our implementation is a ``lookback mode'' that calculates spatially dependent redshifts within the computational volume thus mimicking a redshift survey lightcone as in \cite{kitaura2016}. In this mode we calculate the comoving distance to individual points from the center of the computational volume \cite{hogg1999} and then derive the expected redshift as seen by an observer there. 

The lookback realizations approximate the output of (for example) a series of realizations at a sequence of redshifts. One difference is that the viscosity parameter $\nu$ is, for numerical speed, uniform over space in all cases.  Thus the shock ``capturing'' in the lightcone realizations is stronger at higher redshift. At the resolution of the lightcone realizations here, this feature does not impact the results. 

\bibliographystyle{JHEP}
\bibliography{main}

\end{document}